\documentclass[12pt,onecolumn]{IEEEtran}
	\usepackage{graphicx,amssymb,amsmath}
	\usepackage{multicol}
	\usepackage[noadjust]{cite}
	\usepackage{setspace}
	\usepackage{subfigure}
	\usepackage{graphicx}
	\usepackage{float}
	\usepackage {url}
	\usepackage{stfloats}
	\usepackage{amsthm,pifont}
	\usepackage{flushend}
	\usepackage{cases,subeqnarray}
	\usepackage{bm,multirow,bigstrut}
	\usepackage{amsmath, amsthm, amssymb}
	\usepackage{textcomp}
	\usepackage{latexsym,bm}
	\usepackage{booktabs}
	\usepackage{xcolor}
	\usepackage{mathtools}
	\usepackage{dsfont}
	\usepackage{extarrows}
	\usepackage{epsfig}
	\usepackage{epsfig}
	\usepackage{epstopdf}
	\theoremstyle{plain}

	\theoremstyle{plain}
	\newtheorem{rem}{Remark}
	\newtheorem{them}{Theorem}
	
	\newtheorem{prop}{Proposition}
	\usepackage{amsmath}

	\IEEEoverridecommandlockouts
	
\begin{document}
	\title{Distribution of the Sum of Fisher-Snedecor $\mathcal{F}$ Random Variables and Its Applications}
	\author{Hongyang Du, Jiayi~Zhang,~\IEEEmembership{Member,~IEEE}, Julian Cheng,~\IEEEmembership{Senior~Member,~IEEE},
		
		and Bo Ai,~\IEEEmembership{Senior~Member,~IEEE}
		\thanks{H. Du and J.~Zhang are with the School of Electronic and Information Engineering, Beijing Jiaotong University, Beijing 100044, P. R. China. (e-mail: \{17211140; jiayizhang \}@bjtu.edu.cn)}
		\thanks{J. Cheng is with with the School of Engineering, The University of British Columbia, Kelowna, BC V1V 1V7, Canada.}
		\thanks{B. Ai is with the  State Key Laboratory of Rail Traffic Control and Safety, Beijing Jiaotong University, Beijing 100044, China (e-mail: boai@bjtu.edu.cn).}
	}
	\maketitle
	\begin{abstract}
		The statistical characterization of the sum of random variables (RVs) are useful for investigating the performance of wireless communication systems. We derive exact closed-form expressions for the probability density function (PDF) and cumulative distribution function (CDF) of a sum of independent but not identically distributed (i.n.i.d.) Fisher-Snedecor $\mathcal{F}$ RVs. Both PDF and CDF are given in terms of the multivariate Fox's $H$-function. Besides, a simple and accurate approximation to the sum of i.n.i.d. Fisher-Snedecor $\mathcal{F}$ variates is presented using the moment matching method. The obtained PDF and CDF are used to evaluate the performance of wireless communication applications including the outage probability, the effective capacity and the channel capacities under four different adaptive transmission strtegies. Moreover, the corresponding approximate expressions are obtained to provide useful insights for the design and deployment of wireless communication systems. In addition, we derive simple asymptotic expressions for the proposed mathematical analysis in the high signal-to-noise ratio regime. Finally, the numerical results demonstrate the accuracy of the derived expressions.
		
	\end{abstract}
	\IEEEpeerreviewmaketitle
	\section{Introduction}
		Recently, the Fisher-Snedecor $\mathcal{F}$ distribution was proposed \cite{7886273} as a tractable fading model to describe the combined effects of shadowing and multipath fading.	This distribution can be reduced to some common fading models, such as Nakagami-$m$ and Rayleigh fading channels. Furthermore, it is found in \cite{7886273} that the $\mathcal{F}$ distribution can provide a better fit to the experimental data obtained for device-to-device (D2D) and wearable communication links, especially at 5.8 GHz, as compared with the well established generalized-$K$ (GK) distribution. In addition, its probability density function (PDF) consists of only elementary functions and it leads to more tractable analysis than the GK model \cite{7886273}. Due to its promising properties, the performance of digital  communication systems over $\mathcal{F}$ distributed fading channels has been analyzed in \cite{kong2018physical,kapucu2019analysis,yoo2019entropy,zhao2019ergodic} and the references therein.
	
	The sum of random variables (RVs) has a wide range of important applications in the performance analysis of wireless communication systems. For example, to enhance the quality of the received signal, maximal-ratio combining (MRC) can be deployed at the receiver to maximize the combiner output signal-to-noise ratio (SNR) \cite{simon2005digital}. The system with MRC receiver operating over different fading channels has been extensively studied \cite{chen2018effective,el2018accurate,zhang2015ergodic,zhang2015multivariate,peppas2009error,zheng2019sum}.
	
	
	The PDF and CDF of the sum of Fisher-Snedecor $\mathcal{F}$ RVs has been derived in terms of Lauricella multivariate hypergeometric function \cite{8359199}. However, there results are difficult to be used in the performance analysis of MRC systems over Fisher-Snedecor $\mathcal{F}$ fading channels due to the complex of the Lauricella multivariate hypergeometric function. Moreover, authors in \cite{8359199} obtain outage probability (OP) and outage capacity expressions involving $L$-fold  Mellin-Barnes type contour integral, where $L$ is the number of diversity branches. These results cannot be calculated efficiently, thus the insights into the system performance are limited. Moreover, the statistical characterization is challenging, if not impossible, to derive the important performance metrics, such as channel capacity and effective capacity.
	
	
	In order to fill this gap, we re-investigate the statistical characterization of the statistical characterization of the sum of independent but not identically distributed (i.n.i.d.) Fisher-Snedecor
	$\mathcal{F}$ RVs and leverage the PDF and CDF expressions to analyze the performance of the MRC receiver in terms of outage probability, channel capacity and effective capacity \cite{sanchez2019statistics,8703806,alhennawi2015closed,chergui2018rician}.

	The main contributions of this paper are summarized as follows:
\begin{itemize}
\item We derive exact closed-form expressions for the PDF and cumulative distribution function (CDF) of the sum of i.n.i.d. Fisher-Snedecor $\mathcal{F}$ RVs in terms of the multivariate Fox's $H$-function, which can be efficiently programmed in standard software packages (e.g., Maple and Mathematica) [21-24].
\item We introduce a paradigm based on the moment matching method to obtain a simple approximation to the sum of Fisher-Snedecor $\mathcal{F}$ RVs using another single $\mathcal{F}$ RV. To improve the accuracy of the approximation, we propose an adjustment factor to modify the error in the lower and upper tail regions. The approximated expression is easier to evaluate in the performance analysis. We employ the Kolmogorov-Smirnov (KS) goodness-of-fit	statistical test to show that the single $\mathcal{F}$ distribution is a highly accurate approximation to the sum of $\mathcal{F}$ RVs.
\item We derive novel analytical expressions for important performance metrics, namely the OP, the effective capacity and the channel capacities under four different adaptive transmission schemes, including optimal rate adaptation with constant transmit power (ORA), simultaneous optimal power and rate adaptation (OPRA), channel inversion with fixed rate (CIFR) and truncated channel inversion with fixed rate (TIFR). Moreover, the final value theorem is used to avoid the conflict between the definition of the multivariate Fox's $H$-function and the analytical expressions.

\item We derive highly accurate and simplified closed-form approximations for the studied performance metrics by using a single $\mathcal{F}$ distribution. Furthermore, we pursue an asymptotic performance analysis in the high-SNR regime. The derived results can provide useful insights into the effects of different system and fading parameters on the performance.

\end{itemize}

The remainder of the paper is organized as follows. In Section \ref{Sention2}, we introduce the statistical characterizations of the $\mathcal{F}$ distribution and the definition of multivariable Fox's $H$ function. Exact closed-form PDF and CDF expressions of the sum of Fisher-Snedecor $\mathcal{F}$ RVs are derived in Section \ref{Sention3}. Section \ref{Sention4} provides a single $\mathcal{F}$ distribution to approximate the distribution of sum of $\mathcal{F}$ RVs using the moment matching method for the first, second and third moments, and the KS goodness-of-fit statistical test is evaluated. In Section \ref{Section5}, we investigate the performance in several wireless communications scenarios, and present simple asymptotic expressions. Section \ref{Section6} provides the numerical results and the accuracy of the obtained expressions  is validated via Monte Carlo simulations. Finally, Section \ref{Section7} concludes this paper.
	\section{Preliminaries}\label{Sention2}
	\subsection{Statistics of Fisher-Snedecor $\mathcal{F}$ Random Variables}
	The PDF and CDF of the instantaneous SNR, $\gamma$, at the destination over Fisher-Snedecor $\mathcal{F}$ fading channels are respectively given by \cite[eq. (6)]{yoo2019comprehensive}, \cite[eq. (12)]{yoo2019comprehensive}.
	\begin{equation}
	{f_{_\gamma }}\left( \gamma  \right) = \frac{{{m^m}{{\left( {{m_s} - 1} \right)}^{{m_s}}}{{\bar \gamma }^{{m_s}}}{\gamma ^{m - 1}}}}{{B\left( {m,{m_s}} \right){{\left({m\gamma  + \left( {{m_s} - 1} \right)\bar \gamma } \right)}^{m + {m_s}}}}},
	\end{equation}
	and
	\begin{align}\label{singleF}
	{{F}_{_{\gamma }}}\left( \gamma  \right)=&\frac{{{m}^{m-1}}{{\gamma }^{m}}}{B\left( m,{{m}_{s}} \right){{\left( {{m}_{s}}-1 \right)}^{m}}{{{\bar{\gamma }}}^{m}}} {}_{2}{{F}_{1}}\left( m,m+{{m}_{s}},m+1;-\frac{m\gamma }{\left( {{m}_{s}}-1 \right)\bar{\gamma }} \right)
	\end{align}	
	where $B(\cdot, \cdot)$ denotes the beta function \cite[eq. (8.38)]{gradshteyn2007}; ${}_2{F_1}\left(  \cdot  \right)$ denotes the Gauss hypergeometric function \cite[eq. (9.10)]{gradshteyn2007} and ${m_s} > 1$; the parameters $m$, $m_{s}$, and $\overline{\gamma}$ denote the number of multipath clusters, shadowing shape, and average SNR, respectively

	The MGF of $\gamma$ is given by \cite[eq. (10)]{yoo2019comprehensive}
	\begin{equation}\label{MGF_F}
	{{\cal M}_\gamma }\left( {s} \right) = {}_1{F_1}\left( {m;1 - {m_s};\frac{{s\bar \gamma \left( {{m_s} - 1} \right)}}{m}} \right) + \frac{{\Gamma \left( { - {m_s}} \right){{\left( {\frac{{s\bar \gamma \left( {{m_s} - 1} \right)}}{m}} \right)}^{{m_s}}}}}{{B\left( {m,{m_s}} \right)}}{}_1{F_1}\left( {m + {m_s};1 + {m_s};\frac{{s\bar \gamma \left( {{m_s} - 1} \right)}}{m}} \right)
	\end{equation}
	where $_1 F_{1}(\cdot, \cdot, \cdot)$ denotes the Kummer confluent hypergeometric function \cite[eq. (9.210.1)]{gradshteyn2007}. With the definition of Tricomi’s confluent hypergeometric function \cite[eq. (9.210.2)]{gradshteyn2007} and after some algebriac manipulations, we obtain
	\begin{equation}\label{MGF_F2}
	{{\cal M}_\gamma }\left( s \right) = \frac{{\Gamma \left( {m + {m_s}} \right)}}{{\Gamma \left( {{m_s}} \right)}}\Psi \left( {m,1 - {m_s};\frac{{s\bar \gamma \left( {{m_s} - 1} \right)}}{m}} \right)
	\end{equation}
	where $\Psi \left( { \cdot , \cdot ; \cdot } \right)$ is the Tricomi’s confluent hypergeometric function, which can be expressed as the Meijer ’s $G$-function \cite[eq. (8.4.3.1)]{gradshteyn2007}, and \eqref{MGF_F2} can be written as eq. (2) in \cite{8359199}.
	
	The $n$th moment of the Fisher-Snedecor $\mathcal{F}$ distribution can be derived in closed-form as \cite[eq. (9)]{yoo2019comprehensive}. With the aid of \cite[eq. (8.384.1)]{gradshteyn2007}, one obtains
	\begin{equation}\label{momenteq}
	{\rm{E}}\left[ {{\gamma ^n}} \right] = {\left( {\frac{{\left( {{m_s} - 1} \right)\bar \gamma }}{m}} \right)^n}\frac{{B\left( {m + n,{m_s} - n} \right)}}{{B\left( {m,{m_s}} \right)}}
	\end{equation}
	where ${\rm E}\left[\cdot  \right]$ denotes the mathematical expectation.
	\subsection{Multivariable Fox's H-function}
	Multivariable Fox's $H$-function has several notations. Among them, we choose a widely adopted notation given as \cite[eq. (A.1)]{mathai2009h}
	\begin{align}\label{definitionH}
	H\left[ {{z_1}, \ldots ,{z_r}} \right] \triangleq & H_{p,q:{p_1},{q_1}; \ldots ;{p_r},{q_r}}^{0,n:{m_1},{n_1}; \ldots ;{m_r},{n_r}}\left[ {\left. {\begin{array}{*{20}{c}}
	{{z_1}}\\
	\vdots \\
	{{z_r}}
	\end{array}} \right|\begin{array}{*{20}{c}}
	{{{\left( {{a_j};\alpha _j^{(1)}, \ldots ,\alpha _j^{(r)}} \right)}_{1,p}}:{{\left( {c_j^{(1)},\gamma _j^{(1)}} \right)}_{1,{p_1}}}; \cdots ;{{\left( {c_j^{(r)},\gamma _j^{(r)}} \right)}_{1,{p_r}}}}\\
	{{{\left( {{b_j};\beta _j^{(1)}, \ldots ,\beta _j^{(r)}} \right)}_{1,q}}:{{\left( {d_j^{(1)},\delta _j^{(1)}} \right)}_{1,{q_1}}}; \cdots ;{{\left( {d_j^{(r)},\delta _j^{(r)}} \right)}_{1,{q_r}}}}
	\end{array}} \right]\notag\\
	= &\frac{1}{{{{(2\pi j)}^r}}}\int_{{L_1}}  \cdots  \int_{{L_r}} \Psi  \left( {{\zeta _1}, \ldots ,{\zeta _r}} \right)\left\{ {\prod\limits_{i = 1}^r {{\phi _i}} \left( {{\zeta _i}} \right)z_i^{{\zeta _i}}} \right\}{d}{\zeta _1} \cdots {d}{\zeta _r}
	\end{align}
	where $j \triangleq \sqrt{-1}$,
	\begin{subequations}
	\begin{equation}
	\Psi \left( {{\zeta _1}, \ldots ,{\zeta _r}} \right) = \frac{{\prod\limits_{j = 1}^n \Gamma  \left( {1 - {a_j} + \sum\limits_{i = 1}^r {\alpha _j^{(i)}} {\zeta _i}} \right)}}{{\left[ {\prod\limits_{j = n + 1}^p \Gamma  \left( {{a_j} - \sum\limits_{i = 1}^r {\alpha _j^{(i)}} {\zeta _i}} \right)} \right]\left[ {\prod\limits_{j = 1}^q \Gamma  \left( {1 - {b_j} + \sum\limits_{i = 1}^r {\beta _j^{(i)}} {\zeta _i}} \right)} \right]}},
	\end{equation}
	\begin{equation}
	{\phi _i}\left( {{\zeta _i}} \right) = \frac{{\left[ {\prod\limits_{\lambda  = 1}^{{m_i}} \Gamma  \left( {d_\lambda ^{(i)} - \delta _\lambda ^{(i)}{\zeta _i}} \right)} \right]\left[ {\prod\limits_{j = 1}^{{n_i}} \Gamma  \left( {1 - c_j^{(i)} + \gamma _j^{(i)}{\zeta _i}} \right)} \right]}}{{\left[ {\prod\limits_{j = {n_i} + 1}^{{p_i}} \Gamma  \left( {c_j^{(i)} - \gamma _j^{(i)}{\zeta _i}} \right)} \right]\left[ {\prod\limits_{\lambda  = {m_i} + 1}^{{q_i}} \Gamma  \left( {1 - d_\lambda ^{(i)} + \delta _\lambda ^{(i)}{\zeta _i}} \right)} \right]}}.
	\end{equation}	
	\end{subequations}
	
	Although the numerical evaluation for multivariate Fox’s $H$-function is not available in popular mathematical packages such as MATLAB and Mathematica, its efficient implementations have been reported. For example, two Mathematica implementations of the single Fox’s $H$-function are provided in \cite{sanchez2019statistics} and \cite{8703806}, a Python implementation for the multivariable Fox’s $H$-function is presented in \cite{alhennawi2015closed}, and an efficient GPU-oriented MATLAB routine for the multivariate Fox’s $H$-function is introduced in \cite{chergui2018rician}. In the following, we will utilize these novel implementations to evaluate our results.

	\section{Sum of Fisher-Snedecor $\mathcal{F}$ Random Variables}\label{Sention3}
	In this section, we investigate the statistical characterization
	of the sum of Fisher-Snedecor $\mathcal{F}$ RVs and derive exact closed-form expressions for PDF and CDF. Let $z \triangleq\  {\gamma_1} + {\gamma_2} +  \cdots  + {\gamma_L}$, where $\gamma_{\ell} \sim \mathcal{F}\left(\overline{\gamma}_{\ell}, m_{\ell}, m_{s_{\ell}}\right)$ $(\ell=1, \cdots, L)$ are i.n.i.d. Fisher-Snedecor $\mathcal{F}$ RVs.
	\begin{them}\label{P1}
	The PDF of the sum of Fisher-Snedecor $\mathcal{F}$ RVs $z$ is given by
	\begin{equation}\label{PDFfinal}
	{f_Z}\!\left( z \right) \!=\! \frac{1}{{z\prod\limits_{\ell = 1}^L {\Gamma \left( {{m_\ell}} \right)\Gamma \left( {{m_{{s_\ell}}}} \right)} }}{H_{{\rm PDF}}}
	\end{equation}
	where
	\begin{equation}
	{H_{{\rm PDF}}}\triangleq H_{0,1:2,1;2,1; \cdots ;2,1}^{0,0:1,2;1,2; \cdots ;1,2}\!\!\left( {\left. {\begin{array}{*{20}{c}}
	{\!\frac{{{m_1}z}}{{\left( {{m_{{s_1}}} - 1} \right){{\bar \gamma }_1}}}}\\
	\vdots \\
	{\frac{{{m_L}z}}{{\left( {{m_{{s_L}}} - 1} \right){{\bar \gamma }_L}}}}
	\end{array}} \right|\begin{array}{*{20}{c}}
	{\!\!\! - :\left( {1,1} \right),\left( {1 \!-\! {m_{{s_1}}},1} \right); \!\cdots\! ;\left( {1,1} \right),\left( {1 \!-\! {m_{{s_L}}},1} \right)}\\
	{\left( {1;1, \!\cdots\! ,1} \right):\left( {{m_1},1} \right);\! \cdots\! ;\left( {{m_L},1} \right)}
	\end{array}} \right).
	\end{equation}

	\begin{IEEEproof}
		Please refer to Appendix \ref{AppendixA}.
	\end{IEEEproof}
	
	\end{them}
	\begin{them}
	The CDF of the sum of Fisher-Snedecor $\mathcal{F}$ RVs $z$ is given by
	\begin{equation}\label{CDF}
	{F_Z}\!\left( z \right) \!=\! \frac{1}{{\prod\limits_{\ell = 1}^L {\Gamma \left( {{m_\ell}} \right)\Gamma \left( {{m_{{s_\ell}}}} \right)} }}{H_{{\rm CDF}}}
	\end{equation}
	where
	\begin{equation}
	{H_{{\rm CDF}}}\triangleq H_{0,1:2,1;2,1; \cdots ;2,1}^{0,0:1,2;1,2; \cdots ;1,2}\!\!\left( {\left. {\begin{array}{*{20}{c}}
	{\!\frac{{{m_1}z}}{{\left( {{m_{{s_1}}} - 1} \right){{\bar \gamma }_1}}}}\\
	\vdots \\
	{\frac{{{m_L}z}}{{\left( {{m_{{s_L}}} - 1} \right){{\bar \gamma }_L}}}}
	\end{array}} \right|\begin{array}{*{20}{c}}
	{\! \!- :\left( {1,1} \right),\left( {1 \!-\! {m_{{s_1}}},1} \right); \!\cdots\! ;\left( {1,1} \right),\left( {1 \!-\! {m_{{s_L}}},1} \right)}\\
	{\left( {0;1, \!\cdots\! ,1} \right):\left( {{m_1},1} \right);\! \cdots\! ;\left( {{m_L},1} \right)}
	\end{array}} \right)
	\end{equation}
	\begin{IEEEproof}
	Following similar procedures as in Appendix \ref{AppendixA}, we can derive the CDF of $z$ by taking the inverse Laplace transform of ${M_z}\left( s \right)/s$.
	\end{IEEEproof}
	\end{them}
	\section{Accurate Closed-Form Approximations}\label{Sention4}
	In this section, we present accurate closed-form approximations to the distribution of a sum of Fisher-Snedecor $\mathcal{F}$ RVs using a single Fisher-Snedecor $\mathcal{F}$ RV, which can be used to provide more insights into the impact of the parameters on the overall system performance. The parameters ${\bar \gamma }$, ${m_{\cal F}}$ and ${m_{{s_{\cal F}}}}$ are obtained using the moment matching method for the first, second and third moments.
	
	\subsection{Single $\mathcal{F}$ Approximation to the sum of Squared $\mathcal{F}$-distributed RVs}
	\begin{them}\label{P2}
	For the sum of i.n.i.d. $\mathcal{F}$ RVs, the parameters of single $\mathcal{F}$ distribution are given by
	\begin{equation}\label{MatchParematers}
	\left\{ \begin{array}{l}
	{{\bar \gamma }_{\cal F}}{=}\sum\limits_{i = 1}^L {{{\bar \gamma }_i}} ,\\
	{m_{\cal F}} =  - \frac{{2\left( {{H_{\cal F}} - {Y_{\cal F}}} \right)}}{{{H_{\cal F}} - 2{Y_{\cal F}} + {H_{\cal F}}{Y_{\cal F}}}},\\
	{m_{{s_{\cal F}}}} = \frac{{4{H_{\cal F}} - 3{Y_{\cal F}} - 1}}{{2{H_{\cal F}} - {Y_{\cal F}} - 1}}
	\end{array} \right.
	\end{equation}
	where $H_{\cal F}$ and $Y_{\cal F}$ can be calculated as
	\begin{equation}\label{MomentMatch}
	\left\{ \begin{array}{l}
	{H_{\cal F}} \triangleq \frac{{\sum\limits_{\ell = 1}^L {\left( {{H_\ell} - {\varepsilon _\ell} - 1} \right){{\bar \gamma }_\ell}^2} }}{{{{\left( {\sum\limits_{\ell= 1}^L {{{\bar \gamma }_\ell}} } \right)}^{2}}}} + 1,\\
	{Y_{\cal F}} \triangleq \frac{{\sum\limits_{\ell = 1}^L {\left( {{H_\ell}{Y_\ell} - {\varepsilon _\ell}{Y_\ell} - 1} \right){{\bar \gamma }_\ell}^3}  + {{\left( {\sum\limits_{\ell= 1}^L {{{\bar \gamma }_\ell}} } \right)}^3} + 3\left( {\sum\limits_{\ell= 1}^L {{{\bar \gamma }_\ell}} } \right)\left( {\sum\limits_{\ell= 1}^L {\left( {{H_\ell} - {\varepsilon _\ell} - 1} \right){{\bar \gamma }_\ell}^2} } \right) - 3\sum\limits_{\ell = 1}^L {\left( {{H_\ell} - {\varepsilon _\ell} - 1} \right){{\bar \gamma }_\ell}^3} }}{{\left( {\sum\limits_{\ell = 1}^L {{{\bar \gamma }_\ell}} } \right)\left( {\sum\limits_{\ell = 1}^L {\left( {{H_\ell} - {\varepsilon _\ell} - 1} \right){{\bar \gamma }_\ell}^2} {+}{{\left( {\sum\limits_{\ell = 1}^L {{{\bar \gamma }_\ell}} } \right)}^{2}}} \right)}}
	\end{array} \right.
	\end{equation}
	where ${H_\ell } = \frac{{\left( {1 + {m_\ell }} \right)\left( {{m_{{s_\ell }}} - 1} \right)}}{{{m_\ell }\left( {{m_{{s_\ell }}} - 2} \right)}}$ $(\ell=1, \cdots, L)$, ${Y_\ell } = \frac{{\left( {{m_{{s_\ell }}} - 1} \right)\left( {2 + {m_\ell }} \right)}}{{{m_\ell }\left( {{m_{{s_\ell }}} - 3} \right)}}$ and $\varepsilon_{\ell}$ is the factor that can be adjusted to minimize the difference between the approximate and the exact statistics. For example, we can choose $\varepsilon_{\ell}$ to minimize the Kolmogorov distance.
	
	For the i.i.d case, let ${H_\ell }=H$ $(\ell=1, \cdots, L)$, ${Y_\ell }=Y$, $m=m_{\ell}$, $m_s=m_{s_\ell}$, $\varepsilon=\varepsilon _\ell$, ${\bar \gamma}={\bar \gamma}_\ell$, so the parameters of single $\mathcal{F}$ distribution are given by
	
	\begin{equation}
	\left\{ \begin{array}{l}
	{{\bar \gamma }_{\cal F}}{=}L {{\bar \gamma }} ,\\
	{m_{\cal F}} =  - \frac{{2\left( {{H_{\cal F}} - {Y_{\cal F}}} \right)}}{{{H_{\cal F}} - 2{Y_{\cal F}} + {H_{\cal F}}{Y_{\cal F}}}},\\
	{m_{{s_{\cal F}}}} = \frac{{4{H_{\cal F}} - 3{Y_{\cal F}} - 1}}{{2{H_{\cal F}} - {Y_{\cal F}} - 1}}
	\end{array} \right.
	\end{equation}
	where
	\begin{equation}
	\left\{ \begin{array}{l}
	{H_{\cal F}} \triangleq \frac{{\left( {H - \varepsilon  - 1} \right)}}{L} + 1,\\
	{Y_{\cal F}} \triangleq \frac{{\left( {H - \varepsilon } \right)Y + {L^2} + 3L\left( {H - \varepsilon } \right) - 3L - 3\left( {H - \varepsilon } \right) + 2}}{{L\left( {H - \varepsilon } \right) - L{+}{L^2}}}.
	\end{array} \right.
	\end{equation}

	\begin{IEEEproof}
	Please refer to Appendix \ref{AppendixB}.
	\end{IEEEproof}
	\end{them}
	\subsection{KS Goodness-of-fit Test}
	KS goodness-of-fit statistical test can be used to test the accuracy of the proposed approximations \cite{papoulis2002probability}. The KS test is defined as the largest absolute difference between two CDFs, which can be expressed as
	\begin{equation}
	T \buildrel \Delta \over = \max \left| {{F_Z}(z) - {F_{\hat Z}}(z)} \right|
	\end{equation}
	where ${{F_Z}(z)}$ is the analytical CDF of RV $Z$ and ${{F_{\hat Z}}(z)}$ is the
	empirical CDF of RV ${\hat Z}$.
	
	Let us define $H_{0}$ as the null hypothesis under which the observed data of ${\hat Z}$ belong to the CDF of the approximate distribution ${{F_{Z}}(z)}$. Hypothesis $H_{0}$ is accepted if $T<T_{\max }$. The critical value $T_{\max }=\sqrt{-(1 / 2 v) \ln (\alpha / 2)}$ corresponds to a significance level of $\alpha$ \cite{papoulis2002probability}.
	
\begin{figure}[t]
	\centering
	\includegraphics[scale=1]{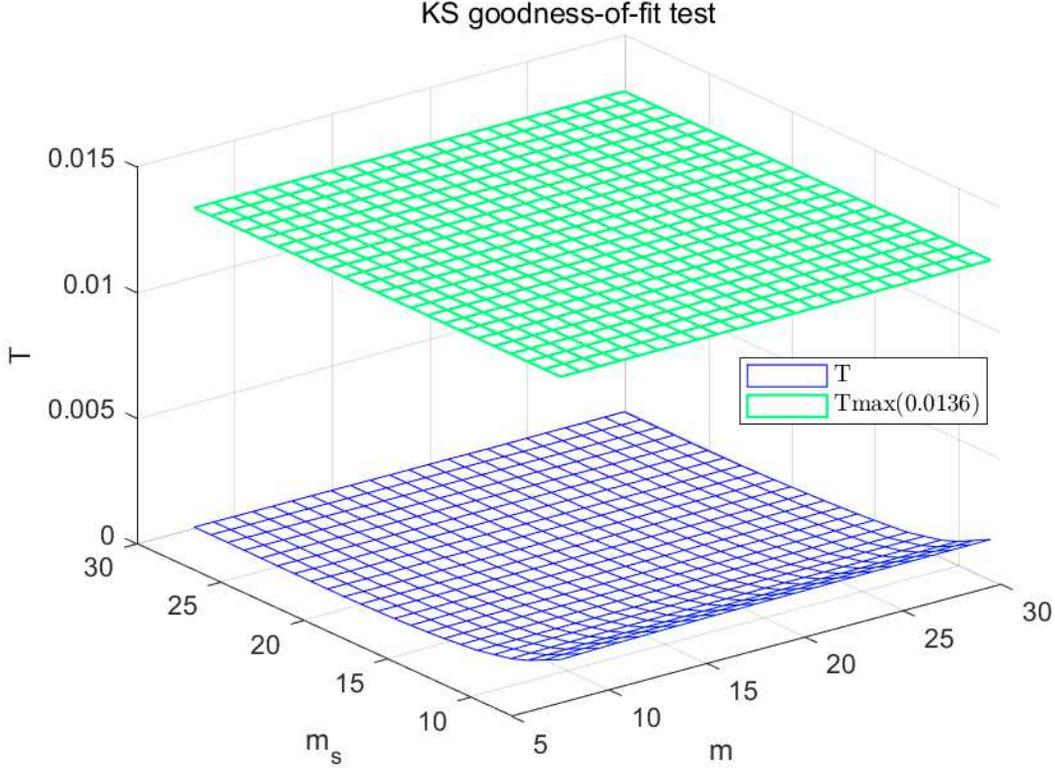}
	\caption{KS goodness-of-fit test statistic for the exact and the approximate	distributions with $5 \%$ significance level for $L=2$.}
	\label{kstestofSUM}
	\end{figure}

\begin{figure}[t]
	\centering
	\includegraphics[scale=1]{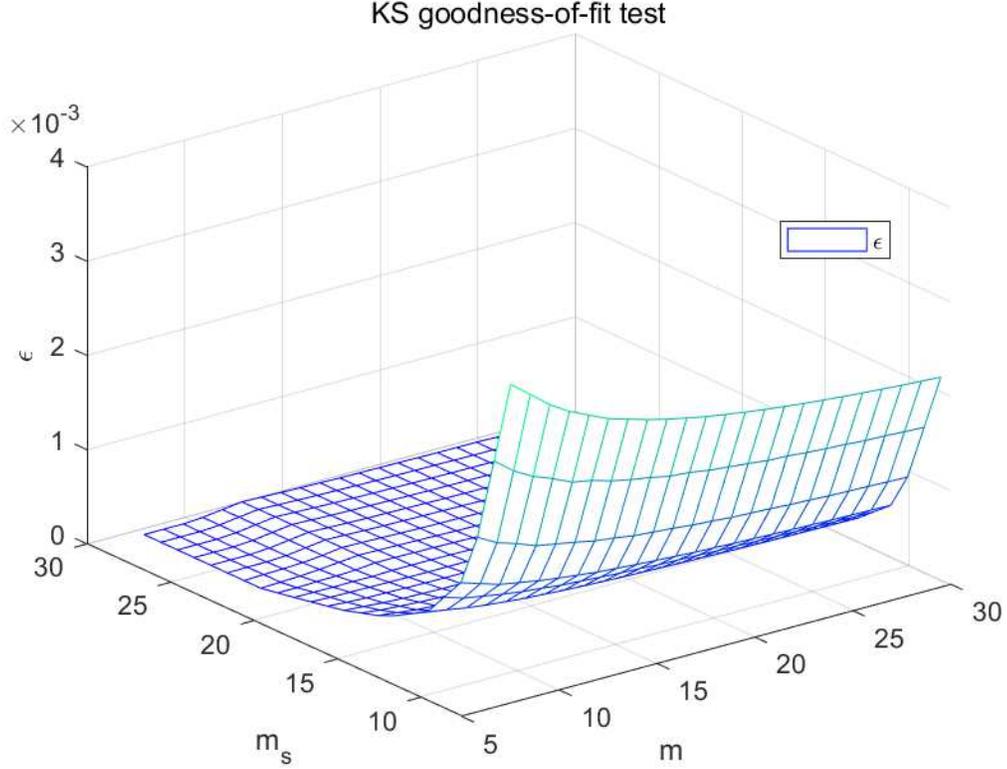}
	\caption{Adjustment factor that minimizes the absolute value of the difference between the exact and the approximate distributions for $L=2$.}
	\label{kstestofSUMyigama}
	\end{figure}

	Without loss of generality, we consider a sum of two i.i.d. Fisher-Snedecor $\mathcal{F}$ RVs with channel parameters $m=m_{\ell}$ and $m_s=m_{s_\ell}$ ($\ell=1, 2$). The average SNR is set by ${\bar \gamma}={\bar \gamma}_\ell=1$ ${\rm{dB}}$. The exact results of CDF have been obtained by averaging at least $v=10^{4}$, and one obtains $T_{\max }=0.0136$ for $\alpha=5\%$. Fig. \ref{kstestofSUM} depicts the KS test statistic for different combinations of $m$ and $m_s$, and the corresponding optimal adjustment factor $\varepsilon$ is shown in the Fig. \ref{kstestofSUMyigama}. It is obvious that $H_{0}$ is accepted with $95 \%$ significance for different settings of parameters. In conclusion, the single $\mathcal{F}$ distribution is a highly accurate approximation to the sum of $\mathcal{F}$ RVs.
	
	\section{Applications to Wireless Communications}\label{Section5}
	In this section, we present six applications in wireless communication systems, including OP, effective capacity, and channel capacities under four different adaptive transmission strategies. We assume communication over a fading channel that follows Fisher-Snedecor $\mathcal{F}$ distribution and a diversity receiver employs MRC.
	
	\subsection{Outage Probability}
	The OP is defined as the probability that the instantaneous
	SNR is less than a predetermined threshold $\gamma _{{\rm{th}}}$. The combiner output SNR $Z$ is simply the sum of the individual branches SNRs and the OP can be directly calculated as
	\begin{equation}
	{P_{{\rm{out}}}} = P\left( {Z < {\gamma _{{\rm{th}}}}} \right){\rm{ = }}F_Z\left( {{\gamma _{{\rm{th}}}}} \right)
	\end{equation}
	where $\gamma _{{\rm{th}}}$ is the minimum usable SNR threshold. Therefore, the OP of the MRC receiver can be directly evaluated by using \eqref{CDF}.
	\begin{prop}
	A highly accurate and simple approximation of OP can be derived using single $\mathcal{F}$ fading channel as
		\begin{equation}
		P_{{\rm{o}},{\cal F}}\simeq\frac{{{m_{\cal F}}^{m_{\cal F}-1}}{{\gamma_{\rm th} }^{m_{\cal F}}}}{B\left( m_{\cal F},{{m}_{s_{\cal F}}} \right){{\left( {{m}_{s_{\cal F}}}-1 \right)}^{m_{\cal F}}}{{{\bar{\gamma_{\cal F}}}}^{m_{\cal F}}}} {}_{2}{{F}_{1}}\left( m_{\cal F},m_{\cal F}+{{m}_{s_{\cal F}}},m_{\cal F}+1;-\frac{m_{\cal F}\gamma_{\rm th} }{\left( {{m}_{s_{\cal F}}}-1 \right)\bar{\gamma }_{\cal F}} \right).
		\end{equation}

	The asymptotic expansions of the OP for high SNRs can be obtained by computing the residue \cite{alhennawi2015closed}. Let us consider the residue at the points $\zeta = \left( {{\zeta_1}, \ldots ,{\zeta_L}} \right)$, where ${\zeta_\ell } = {\min _{j = 1, \ldots ,{m_\ell }}}\left( {d_j^{(\ell )}/\delta_j^{(\ell )}} \right)$ $(\ell=1, \cdots, L)$. We obtain the approximate OP expression as
	\begin{equation}
P_{{\rm{o}},{\rm{appr}}}\simeq\frac{1}{{\Gamma \left( {1 + \sum\limits_{\ell  = 1}^L {{m_\ell }} } \right)}}\prod\limits_{\ell  = 1}^L {\frac{{\Gamma \left( {{m_{{s_\ell }}} + {m_\ell }} \right)}}{{\Gamma \left( {{m_{{s_\ell }}}} \right)}}} {\left( {\frac{{\gamma_{\rm th} {m_\ell }}}{{{{\bar \gamma }_\ell }\left( {{m_{{s_\ell }}} - 1} \right)}}} \right)^{{m_\ell }}}.
	\end{equation}
	\end{prop}
	\subsection{Effective Capacity}
	The effective capacity, which accounts for the achievable capacity subject to the incurred latency relating to the corresponding buffer occupancy, is a useful and insightful information theoretic measure particularly in emerging technologies. The effective capacity normalised by the bandwidth can be defined as \cite[eq. (11)]{gursoy2009analysis}
	\begin{equation}\label{eff}
	{C_{\rm{eff}}} =  - \frac{1}{A}{\log _2}\left( {\int_0^\infty  {\frac{1}{{{{(1 + z)}^A}}}{f_z}\left( z \right)dz} } \right)\quad {\rm{ bit }}/{\rm{s}}/{\rm{Hz}}
	\end{equation}
	where $z=\sum\limits_{\ell  = 1}^L {{\gamma _\ell }}$, $\gamma_\ell \sim \mathcal{F}\left(\overline{\gamma}_\ell, m_\ell, m_{{s}_\ell}\right)$ $(\ell=1,\cdots,2)$, and
	\begin{equation}
	A = \frac{{BT\theta }}{{\ln 2}}
	\end{equation}
	is a delay constraint with the asymptotic decay rate of the buffer occupancy $\theta$, system bandwidth $B$ and the block length $T$.
	\begin{prop}
	For the i.n.i.d. case, the effective capacity can be deduced as
	\begin{equation}\label{efffinal}
	{C_{\rm{eff}}} =  - \frac{1}{A}{\log _2}\left( {\frac{A}{{\Gamma \left( {1 + A} \right)}}\prod\limits_{\ell  = 1}^L {\frac{{1}}{{\Gamma \left( {{m_\ell }} \right)\Gamma \left( {{m_{{s_\ell }}}} \right)}}} {H_{\rm{eff}}}} \right),
	\end{equation}
	where
	\begin{equation}
	{H_{\rm{eff}}}\!\triangleq \! H_{1,0:2,1; \cdots ;2,1}^{0,1:1,2; \cdots ;1,2}\left(\!\! {\left. {\begin{array}{*{20}{c}}
	{\frac{{{m_1}}}{{\left( {{m_{{s_1}}} - 1} \right){{\bar \gamma }_1}}}}\\
	\vdots \\
	{\frac{{{m_L}}}{{\left( {{m_{{s_L}}} - 1} \right){{\bar \gamma }_L}}}}
	\end{array}} \right|\begin{array}{*{20}{c}}
	{\left( {1\!-\!A; - 1,\!\cdots\!, - 1} \right):\left( {1,1} \right),\left( {1\!-\!{m_{{s_1}}},1} \right); \cdots ;\left( {1,1} \right),\left( {1\!-\!{m_{{s_L}}},1} \right)}\\
	{ - :\left( {{m_1},1} \right); \cdots ;\left( {{m_L},1} \right)}
	\end{array}}\!\! \right),
	\end{equation}
	\end{prop}
	\begin{IEEEproof}
	Using the classical Newton-Leibniz formula, we can rewrite \eqref{eff} as
	\begin{align}\label{effre}
	{C_{\rm{eff}}} =&  - \frac{1}{A}{\log _2}\left( {\int_0^\infty  {\frac{1}{{{{(1 + z)}^A}}}{f_z}\left( z \right)dz} } \right){=} - \frac{1}{A}{\log _2}\left( {\int_0^\infty  {\frac{1}{{{{\left( {1 + z} \right)}^A}}}d{F_Z}\left( z \right)} } \right)\notag\\
	{=}& - \frac{1}{A}{\log _2}\left( {A\underbrace {\int_0^\infty  {\frac{{{F_z }\left( z  \right)}}{{{{\left( {1 + z } \right)}^{A + 1}}}}} dz}_{{I_1}}} \right).
	\end{align}
	With the help of \eqref{CDF}, ${I_1}$ can be expressed as
	\begin{align}
	{I_1} =&\prod\limits_{\ell  = 1}^L {\frac{1}{{\Gamma \left( {{m_\ell }} \right)\Gamma \left( {{m_{{s_\ell }}}} \right)}}} {\left( {\frac{1}{{2\pi j}}} \right)^L}\notag\\
	& \times \int_0^\infty  {\frac{{1}}{{{{\left( {1 + z} \right)}^{A + 1}}}}\int_{{{\cal L}_1}} {\int_{{{\cal L}_2}} { \cdots \int_{{{\cal L}_{_L}}} {\frac{1}{{\Gamma \left( {1 + \sum\limits_{\ell  = 1}^L {{\zeta _\ell }} } \right)}}} \left\{ {\prod\limits_{\ell  = 1}^L {\Upsilon \left( {{\zeta _\ell }} \right)} {{\left( {\frac{{z{m_\ell }}}{{{{\bar \gamma }_\ell }\left( {{m_{{s_\ell }}} - 1} \right)}}} \right)}^{ - {\zeta _\ell }}}} \right\}} d{\zeta _1}d{\zeta _2} \cdots d{\zeta _L}} } dz
	\end{align}
	where $\Upsilon \left( {{\zeta _\ell }} \right) = \Gamma \left( {{m_\ell } - {\zeta _\ell }} \right)\Gamma \left( {{\zeta _\ell }} \right)\Gamma \left( {{m_{{s_\ell }}} + {\zeta _\ell }} \right)$. According to Fubini's theorem, we can exchange the order of integrations in ${I_1}$, and derive
	\begin{align}\label{effjifen1}
	{I_1} =& \prod\limits_{\ell  = 1}^L {\frac{1}{{\Gamma \left( {{m_\ell }} \right)\Gamma \left( {{m_{{s_\ell }}}} \right)}}} {\left( {\frac{1}{{2\pi j}}} \right)^L}\notag\\
	&\times \int_{{{\cal L}_1}} {\int_{{{\cal L}_2}} { \cdots \int_{{{\cal L}_{_L}}} {\frac{1}{{\Gamma \left( {1 + \sum\limits_{\ell  = 1}^L {{\zeta _\ell }} } \right)}}} \left\{ {\prod\limits_{\ell  = 1}^L {\Upsilon \left( {{\zeta _\ell }} \right)} {{\left( {\frac{{{m_\ell }}}{{{{\bar \gamma }_\ell }\left( {{m_{{s_\ell }}} - 1} \right)}}} \right)}^{ - {\zeta _\ell }}}} \right\}\underbrace {\int_0^\infty  {\frac{{{z^{\sum\limits_{\ell  = 1}^L {{\zeta _\ell }} }}}}{{{{\left( {1 + z} \right)}^{A + 1}}}}} dz}_{{I_{2}}}} d{\zeta _1}d{\zeta _2} \cdots d{\zeta _L}}.
	\end{align}
	Using \cite[eq. (3.194.3)]{gradshteyn2007}, we can slove $I_2$ as
	\begin{equation}\label{effI2}
	{I_{2}} = \frac{{\Gamma \left( {\sum\limits_{\ell  = 1}^L {{\zeta _\ell }}  + 1} \right)\Gamma \left( {A - \sum\limits_{\ell  = 1}^L {{\zeta _\ell }} } \right)}}{{\Gamma \left( {1 + A} \right)}}.
	\end{equation}
	Combining \eqref{effI2}, \eqref{effjifen1}, \eqref{effre} and \eqref{definitionH}, we obtain \eqref{efffinal} to complete the proof.
	\end{IEEEproof}
	\begin{rem}
	A highly accurate approximation of effective capacity can be derived using a single $\mathcal{F}$ fading channel by setting $L=1$ in \eqref{efffinal}. After some algebraic manipulations, we obtain the same result as \cite[eq. (33)]{yoo2019comprehensive}
	which provides useful insights because it can be used as a benchmark for the derivation of simpler approximations or bounds.
	The approximation of effective capacity in the high SNR region under $\mathcal{F}$ fading channels was also derived as \cite[eq. (41)]{yoo2019comprehensive}, which can be used as the approximation of \eqref{efffinal}.
	
	\end{rem}
	\subsection{Channel Capacity}
	Channel capacity is a key performance metric in communication systems. In the following, we analyze the channel capacity performance under four diffrernt adaptive transmission schemes, namely CIFR, TIFR, ORA and OPRA.
	\subsubsection{Channel Inversion with Fixed Rate}\label{zhongzhicifr}
	CIFR ensures a fixed data rate at the receiver by inverting the channel and adapting the transmit power. The channel capacity under CIFR is defined in \cite{alouini1999capacity} as
	\begin{equation}\label{cifr}
	{C_{{\rm{CIFR}}}} = B{\log _2}\left( {1 + \frac{1}{{\int_0^\infty  {\frac{{{f_Z}\left( z \right)}}{z}dz} }}} \right).
	\end{equation}

	\begin{prop}
	The channel capacity under CIFR can be expressed as
	\begin{equation}\label{cifrfinal}
	{C_{{\rm{CIFR}}}} = B{\log _2}\left( {1 + \frac{1}{{\prod\limits_{\ell  = 1}^L {\frac{s}{{\Gamma \left( {{m_{{s_\ell }}}} \right)\Gamma \left( {{m_\ell }} \right)}}} {H_{{\rm{CIFR}}}}}}} \right)
	\end{equation}
	where
	\begin{equation}
	{H_{{\rm{CIFR}}}}\triangleq H_{1,0:2,1; \cdots ;2,1}^{0,1:1,2; \cdots ;1,2}\left(\!\!\! {\left. {\begin{array}{*{20}{c}}
	{\frac{{{m_1}}}{{s\left( {{m_{{s_1}}} - 1} \right){{\bar \gamma }_1}}}}\\
	\vdots \\
	{\frac{{{m_L}}}{{s\left( {{m_{{s_L}}} - 1} \right){{\bar \gamma }_L}}}}
	\end{array}} \right|\!\!\!\begin{array}{*{20}{c}}
	{\left( {2;1, \cdots ,1} \right):\left( {1,1} \right),\left( {1 - {m_{{s_1}}},1} \right); \cdots ;\left( {1,1} \right),\left( {1 - {m_{{s_L}}},1} \right)}\\
	{\left( {1;1, \cdots ,1} \right):\left( {{m_1},1} \right); \cdots ;\left( {{m_L},1} \right)}
	\end{array}}\!\!\!\! \right)
	\end{equation}
	and $s$ is a number close to zero (e.g., $s=10^{-6}$).
	\end{prop}
	\begin{IEEEproof}\label{zhongzhiproof}
	With the aid of \eqref{PDFfinal}, the integral in \eqref{cifr} can be written as
	\begin{align}\label{cifrjifen}
	{I_3} =& \prod\limits_{\ell  = 1}^L {\frac{1}{{\Gamma \left( {{m_{{s_\ell }}}} \right)\Gamma \left( {{m_\ell }} \right)}}{{\left( {\frac{1}{{2\pi i}}} \right)}^L}} \notag\\
	&\times \int_0^\infty  {\int_{{{\cal L}_1}} {\int_{{{\cal L}_2}} { \cdots \int_{{{\cal L}_{_L}}} {\prod\limits_{\ell  = 1}^L {\Upsilon \left( {{\zeta _\ell }} \right)} } {{\left( {\frac{{{m_\ell }}}{{{{\bar \gamma }_\ell }\left( {{m_{{s_\ell }}} - 1} \right)}}} \right)}^{{\zeta _\ell }}}} \frac{1}{{\Gamma \left( {\sum\limits_{\ell  = 1}^L {{\zeta _\ell }} } \right)}}{z^{ - 2 + \sum\limits_{\ell  = 1}^L {{\zeta _\ell }} }}{d{\zeta _1}d{\zeta _2} \cdots d{\zeta _L}}} dz}.
	\end{align}
Let ${\cal L}\left\{ {p\left( t \right)} \right\} = P\left( x \right)$. According to the property of Laplace transform, we have
	\begin{equation}
	{\cal L}\left\{ {\int_0^t {p\left( z \right)dz} } \right\} = \frac{{P\left( x \right)}}{x}.
	\end{equation}
	With the help of the final value theorem, it follows that
	\begin{equation}\label{finaltheorem}
	\mathop {\lim }\limits_{t \to \infty } \left( {\int_0^t {p\left( z \right)dz} } \right) = s\frac{{P\left( s \right)}}{s} = P\left( s \right).
	\end{equation}
We can use \eqref{finaltheorem} to obtain
	\begin{equation}\label{cifrjifen2}
	\int_0^\infty  {{z^{ - 2 + \sum\limits_{\ell  = 1}^L {{\zeta _\ell }} }}dz}  = {\cal L}\left\{ {{z^{ - 2 + \sum\limits_{\ell  = 1}^L {{\zeta _\ell }} }}} \right\}{=}{\left( {\frac{1}{s}} \right)^{\sum\limits_{\ell  = 1}^L {{\zeta _\ell }}  - 1}}\Gamma \left( {\sum\limits_{\ell  = 1}^L {{\zeta _\ell }}  - 1} \right).
	\end{equation}
	Employing \eqref{cifrjifen} and \eqref{cifrjifen2}, we can rewrite ${I_3}$ as
	\begin{equation}\label{icifr}
	{I_3} = \prod\limits_{\ell  = 1}^L {\frac{s}{{\Gamma \left( {{m_{{s_\ell }}}} \right)\Gamma \left( {{m_\ell }} \right)}}{{\left( {\frac{1}{{2\pi i}}} \right)}^L}} \int_{{{\cal L}_1}} \!{\int_{{{\cal L}_2}} {\!\! \cdots\!\! \int_{{{\cal L}_{_L}}} {\frac{{\Gamma \left( {\sum\limits_{\ell  = 1}^L {{\zeta _\ell }}  - 1} \right)}}{{\Gamma \left( {\sum\limits_{\ell  = 1}^L {{\zeta _\ell }} } \right)}}\prod\limits_{\ell  = 1}^L {\Upsilon \left( {{\zeta _\ell }} \right)} } {{\left( {\frac{{{m_\ell }}}{{s{{\bar \gamma }_\ell }\left( {{m_{{s_\ell }}} - 1} \right)}}} \right)}^{{\zeta _\ell }}}} d{\zeta _1}d{\zeta _2} \cdots d{\zeta _L}}.
	\end{equation}
	Substituting \eqref{icifr} into \eqref{cifr} and using \eqref{definitionH}, we obtain \eqref{cifrfinal} which completes the proof.
	\end{IEEEproof}
	\begin{rem}
	With the aid of Theorem \ref{P2}, a highly accurate approximation of the channel capacity per unit bandwidth under CIFR can be deduced using single $\mathcal{F}$ fading channel. Setting $L=1$ in \eqref{cifrfinal} and we get the same result as \cite[eq. (23)]{zhao2019ergodic} after some algebraic manipulations.
	Notice that \cite[eq. (23)]{zhao2019ergodic} is insightful and we can observe that $m$ and $m_s$ have the same influence to the channel capacity under CIFR approximately. Besides, it is easy to see that the channel capacity under CIFR increases as $m$ and $m_s$ increase.
	Under $\mathcal{F}$ composite fading channel, a high SNR approximation of channel capacity under CIFR was derived as \cite[eq. (26)]{zhao2019ergodic}, which can be used as the high-SNR approximation of \eqref{cifrfinal}.			
	\end{rem}
	\subsubsection{Truncated Channel Inversion with Fixed Rate}
	Another approach is to use a modified inversion policy which inverts the channel fading only above a fixed cutoff fade depth. The channel capacity under TIFR policy is defined as
	\begin{equation}\label{tifr}
	C_{\rm{TIFR}}=B{\log _2}\left( {1 + \frac{1}{{\int_{{z_0}}^\infty  {\frac{{{f_Z}\left( z \right)}}{z}dz} }}} \right)\int_{{z_0}}^\infty  {{f_Z}\left( z \right)dz}
	\end{equation}
	where $z_0$ is a cutoff level that can be selected to achieve a specified outage probability or to maximize \eqref{tifr}. Notice that \eqref{tifr} reduces to \eqref{cifr} when $z_0$ approaches zero.
	\begin{prop}
	The closed-form expression for the capacity under TIFR is given by
	\begin{equation}\label{tifrfinal}
	C_{\rm{TIFR}}= B{\log _2}\left( {1 + \frac{1}{{\frac{1}{{{z_0}}}\prod\limits_{\ell  = 1}^L {\frac{1}{{\Gamma \left( {{m_{{s_\ell }}}} \right)\Gamma \left( {{m_\ell }} \right)}}} {H_{{\rm{TIFR}}}}}}} \right)\left( {1 - \prod\limits_{\ell  = 1}^L {\frac{1}{{\Gamma \left( {{m_{{s_\ell }}}} \right)\Gamma \left( {{m_\ell }} \right)}}} {H_{{\rm{CDF}}}}} \right)
	\end{equation}
	where
	\begin{equation}
	H_{\rm{TIFR}}\! \triangleq\! H_{2,1:2,1; \cdots ;2,1}^{0,1:1,2; \cdots ;1,2}\left(\!\!\! {\left. {\begin{array}{*{20}{c}}
	{\frac{{{z_0}{m_1}}}{{\left( {{m_{{s_1}}} - 1} \right){{\bar \gamma }_1}}}}\\
	\vdots \\
	{\frac{{{z_0}{m_L}}}{{\left( {{m_{{s_L}}} - 1} \right){{\bar \gamma }_L}}}}
	\end{array}}\!\! \right|\!\!\begin{array}{*{20}{c}}
	{(0; - 1,\! \cdots \!, - 1)\!:\!\left( {1,1} \right),\left( {1 - {m_{{s_1}}},1} \right);\! \cdots \!;\left( {1,1} \right),\left( {1 - {m_{{s_L}}},1} \right)}\\
	{\left( { - 1; - 1, \cdots , - 1} \right):\left( {{m_1},1} \right); \cdots ;\left( {{m_L},1} \right)}
	\end{array}}\!\!\! \right).
	\end{equation}
	\end{prop}
	\begin{IEEEproof}
	Observe that
	\begin{equation}\label{ccdftocdf}
	\int_{{z_0}}^\infty  {{f_Z}\left( z \right)dz}  = 1 - \int_0^{{z_0}} {{f_Z}\left( z \right)dz}  = 1 - {F_Z}\left( {{z_0}} \right)
	\end{equation}
	where ${F_Z}\left( {{z_0}} \right)$ can be deduced using \eqref{CDF}. Thus, $C_{\rm{TIFR}}$ can be expressed as
	\begin{equation}\label{tifrjisuan}
	C_{\rm{TIFR}}= B{\log _2}\left( {1 + \frac{1}{{\underbrace {\int_{{z_0}}^\infty  {\frac{{{f_Z}\left( z \right)}}{z}dz} }_{{I_5}}}}} \right)\left( {1 - \prod\limits_{\ell  = 1}^L {\frac{1}{{\Gamma \left( {{m_{{s_\ell }}}} \right)\Gamma \left( {{m_\ell }} \right)}}} {H_{CDF}}} \right).
	\end{equation}
Substituting \eqref{PDFfinal} into \eqref{tifrjisuan}, we can rewrite $I_5$ as
	\begin{align}
	I_5 =& \prod\limits_{\ell  = 1}^L {\frac{1}{{\Gamma \left( {{m_{{s_\ell }}}} \right)\Gamma \left( {{m_\ell }} \right)}}{{\left( {\frac{1}{{2\pi i}}} \right)}^L}}\notag\\
	&\times \int_{{z_0}}^\infty  {\int_{{{\cal L}_1}} {\int_{{{\cal L}_2}} { \cdots \int_{{{\cal L}_{_L}}} {\frac{1}{{\Gamma \left( {\sum\limits_{\ell  = 1}^L {{\zeta _\ell }} } \right)}}\prod\limits_{\ell  = 1}^L {\Upsilon \left( {{\zeta _\ell }} \right)} } {{\left( {\frac{{{m_\ell }}}{{{{\bar \gamma }_\ell }\left( {{m_{{s_\ell }}} - 1} \right)}}} \right)}^{{\zeta _\ell }}}} {z^{ - 2 + \sum\limits_{\ell  = 1}^L {{\zeta _\ell }} }}d{\zeta _1}d{\zeta _2} \cdots d{\zeta _L}} dz}.
	\end{align}
	Note that the order of integration can be interchangeable according to Fubini's theorem, we can express $I_5$ as
	\begin{align}\label{tifrfinaljifen}
	I_5 =& \prod\limits_{\ell  = 1}^L {\frac{1}{{\Gamma \left( {{m_{{s_\ell }}}} \right)\Gamma \left( {{m_\ell }} \right)}}{{\left( {\frac{1}{{2\pi i}}} \right)}^L}}\notag\\
	&\times \int_{{{\cal L}_1}} {\int_{{{\cal L}_2}} { \cdots \int_{{{\cal L}_{_L}}} {\frac{{\Gamma \left( {1 - \sum\limits_{\ell  = 1}^L {{\zeta _\ell }} } \right)}}{{\Gamma \left( {\sum\limits_{\ell  = 1}^L {{\zeta _\ell }} } \right)\Gamma \left( {2 - \sum\limits_{\ell  = 1}^L {{\zeta _\ell }} } \right)}}\prod\limits_{\ell  = 1}^L {\Upsilon \left( {{\zeta _\ell }} \right)} } {{\left( {\frac{{{\gamma _0}{m_\ell }}}{{{{\bar \gamma }_\ell }\left( {{m_{{s_\ell }}} - 1} \right)}}} \right)}^{{\zeta _\ell }}}} d{\zeta _1}d{\zeta _2} \cdots d{\zeta _L}}.
	\end{align}
	Employing \eqref{definitionH} and substituting \eqref{tifrfinaljifen} into \eqref{tifrjisuan}, we obtain \eqref{tifrfinal} to complete the proof.
	\end{IEEEproof}
	\begin{rem}
	Using Theorem \ref{P2}, an accurate approximation of the channel capacity under TIFR can be derived by single $\mathcal{F}$ fading channel. Substituting $L=1$ in \eqref{tifrfinal}. After some algebraic manipulations, we can get the same result as \cite[eq. (27)]{zhao2019ergodic} and \cite[eq. (28)]{zhao2019ergodic}.
	An approximation of channel capacity under TIFR in the high SNR region is given by \cite[eq. (30)]{zhao2019ergodic}, which can be used as the high SNR approximation of \eqref{tifrfinal} with the aid of \eqref{MatchParematers}.
	Furthermore, \cite[eq. (30)]{zhao2019ergodic} is particularly insightful because it only has element functions.
	\end{rem}
	\subsubsection{Optimal Rate Adaptation with Constant Transmit Power}
	The channel	capacity under ORA with a constant transmit power is given by \cite[eq. (29)]{alouini1999capacity}
	\begin{equation}\label{orajifen}
	{C_{\rm{ORA}}} = B\int_0^\infty  {{{\log }_2}\left( {1 + z} \right)} {f_Z}\left( z \right)dz.
	\end{equation}
	\begin{prop}
	Under the ORA scheme, the channel capacity can be expressed as
	\begin{equation}\label{orafinal}
	{C_{\rm{ORA}}}=\frac{B}{{s\ln \left( 2 \right)}}\prod\limits_{\ell  = 1}^L {\frac{1}{{\Gamma \left( {{m_{{s_\ell }}}} \right)\Gamma \left( {{m_\ell }} \right)}}} {H_{ORA}}
	\end{equation}
	where

		\begin{align}
	{H_{\rm{ORA}}} \triangleq H_{4,1:2,1; \cdots ;2,1;1,2}^{0,3:1,2; \cdots ;1,2;1,1}\Bigg({\Delta _{\rm{ORA}}}&\left|{\begin{array}{*{20}{c}}
		{\left( {0; - 1, \cdots , - 1} \right),\left( {0; - 1, \cdots , - 1} \right),\left( {1;1, \cdots ,1} \right),\left( {2;1, \cdots ,1} \right):}\\
		{\left( {1;1, \cdots ,1,0} \right):}
		\end{array}} \right.\notag\\
		& \begin{array}{*{20}{c}}
		{\left( {1,1} \right),\left( {1 - {m_{{s_1}}},1} \right); \cdots ;\left( {1,1} \right),\left( {1 - {m_{{s_L}}},1} \right);\left( {1,1} \right)}\\
		{\left( {{m_1},1} \right); \cdots ;\left( {{m_L},1} \right);\left( {1,1} \right),\left( {0,1} \right)}
		\end{array}\Bigg)
  		\end{align}
and ${\Delta _{\rm{ORA}}} \triangleq {\left( {\frac{{{m_1}}}{{\left( {{m_{{s_1}}} - 1} \right){{\bar \gamma }_1}}}, \cdots ,\frac{{{m_L}}}{{\left( {{m_{{s_L}}} - 1} \right){{\bar \gamma }_L}}},s} \right)^{\rm T}}.$
	
	\end{prop}
	\begin{IEEEproof}
	Substituting \eqref{PDFfinal} into \eqref{orajifen} and changing the order of integration, the channel capacity under ORA can be expressed as
	\begin{align}\label{orachushi}
	{C_{\rm{ORA}}} = & B\prod\limits_{\ell  = 1}^L {\frac{1}{{\Gamma \left( {{m_{{s_\ell }}}} \right)\Gamma \left( {{m_\ell }} \right)}}} {\left( {\frac{1}{{2\pi i}}} \right)^L}\notag\\
	&\times \int_{{{\cal L}_1}} {\int_{{{\cal L}_2}} { \cdots \int_{{{\cal L}_{_L}}} {\frac{1}{{\Gamma \left( {\sum\limits_{\ell  = 1}^L {{\zeta _\ell }} } \right)}}} \left\{ {\prod\limits_{\ell  = 1}^L {\Upsilon \left( {{\zeta _\ell }} \right)} {{\left( {\frac{{{m_\ell }}}{{{{\bar \gamma }_\ell }\left( {{m_{{s_\ell }}} - 1} \right)}}} \right)}^{{\zeta _\ell }}}} \right\}}}\notag\\
	&\times \underbrace {\int_0^\infty  {{{\log }_2}\left( {1 + z} \right)} {z^{ - 1 + \sum\limits_{\ell  = 1}^L {{\zeta _\ell }} }}dz}_{{I_6}}d{\zeta _1}d{\zeta _2} \cdots d{\zeta _L}.
	\end{align}
	With the help of \cite[eq. 2.6.9.21]{Prudnikov1986Integrals} and \cite[eq. 8.334.3]{gradshteyn2007}, $I_6$ can be deduced as
	\begin{equation}\label{L2ora}
	{I_{6}}{=}\int_0^\infty  {{{\log }_2}\left( {1 + z} \right)} {z^{ - 1 + \sum\limits_{\ell  = 1}^L {{\zeta _\ell }} }}d\gamma {=} - \frac{1}{{\ln 2}}\Gamma \left( {\sum\limits_{\ell  = 1}^L {{\zeta _\ell }} } \right)\Gamma \left( { - \sum\limits_{\ell  = 1}^L {{\zeta _\ell }} } \right)
	\end{equation}
	where $- 1 < \mathbb{R}\left( {\sum\limits_{\ell  = 1}^L {{\zeta _\ell }} } \right) < 0$.
	
	In order to avoid conflicts with the defination of the multivariate $H$-function, employing \cite[eq. 2.5.16]{kilbas2004h}, we can express $I_6$ as
	\begin{equation}
	{I_{6}}{=}\frac{1}{{\ln \left( 2 \right)}}\int_0^\infty  {G_{2,2}^{1,2}\left( {z\left| {\begin{array}{*{20}{c}}
	{\sum\limits_{\ell  = 1}^L {{\zeta _\ell }} ,\sum\limits_{\ell  = 1}^L {{\zeta _\ell }} }\\
	{\sum\limits_{\ell  = 1}^L {{\zeta _\ell }} ,\sum\limits_{\ell  = 1}^L {{\zeta _\ell }}  - 1}
	\end{array}} \right.} \right)} dz,
	\end{equation}
	 Using \cite[eq. (9.301)]{gradshteyn2007} and the Laplace transform of Meijer’s $G$-function \cite[eq. 07.34.22.0003.01]{web}, we obtain
	
	\begin{align}
{\cal L}\left\{ {G_{2,2}^{1,2}\left( {z\left| {\begin{array}{*{20}{c}}
			{\sum\limits_{\ell  = 1}^L {{\zeta _\ell }} ,\sum\limits_{\ell  = 1}^L {{\zeta _\ell }} }\\
			{\sum\limits_{\ell  = 1}^L {{\zeta _\ell }} ,\sum\limits_{\ell  = 1}^L {{\zeta _\ell }}  - 1}
			\end{array}} \right.} \right)} \right\}&{=}\frac{1}{x}G_{{3},2}^{1,{3}}\left( {\left. {\frac{{1}}{x}} \right|\begin{array}{*{20}{c}}
	{{0}\sum\limits_{\ell  = 1}^L {{\zeta _\ell }} ,\sum\limits_{\ell  = 1}^L {{\zeta _\ell }} }\\
	{\sum\limits_{\ell  = 1}^L {{\zeta _\ell }} ,\sum\limits_{\ell  = 1}^L {{\zeta _\ell }}  - 1}
	\end{array}} \right)\notag\\
& =\int_{{\cal L}_{L+{1}}} {\frac{{\Gamma \left( {1 - {\zeta _{L + 1}}} \right){\Gamma ^2}\left( {1 - \sum\limits_{\ell  = 1}^{L + 1} {{\zeta _\ell }} } \right)\Gamma \left( {\sum\limits_{\ell  = 1}^{L + 1} {{\zeta _\ell }} } \right)}}{{\Gamma \left( {2 - \sum\limits_{\ell  = 1}^{L + 1} {{\zeta _\ell }} } \right)}}{x^{{\zeta _{L + 1}}}}} d{\zeta _{L + 1}}.
	\end{align}
	Thus, using the final value theorem and following the similar procedures as the proof of \eqref{cifrfinal}, we can rewrite the channel capacity under ORA as
	\begin{align}\label{zuihouora}
	C_{\rm{ORA}}=& \frac{B}{{s\ln \left( 2 \right)}}\prod\limits_{\ell  = 1}^L {\frac{1}{{\Gamma \left( {{m_{{s_\ell }}}} \right)\Gamma \left( {{m_\ell }} \right)}}} {\left( {\frac{1}{{2\pi i}}} \right)^{L + 1}}\notag\\
	&\times \int_{{{\cal L}_1}} {\int_{{{\cal L}_2}} { \cdots \int_{{{\cal L}_{_{L + 1}}}} {\frac{{{\Gamma ^2}\left( {1 - \sum\limits_{\ell  = 1}^{L + 1} {{\zeta _\ell }} } \right)\Gamma \left( {\sum\limits_{\ell  = 1}^{L + 1} {{\zeta _\ell }} } \right)}}{{\Gamma \left( {2 - \sum\limits_{\ell  = 1}^{L + 1} {{\zeta _\ell }} } \right)\Gamma \left( {\sum\limits_{\ell  = 1}^L {{\zeta _\ell }} } \right)}}} \left\{ {\prod\limits_{\ell  = 1}^L {\Upsilon \left( {{\zeta _\ell }} \right)} {{\left( {\frac{{{m_\ell }}}{{{{\bar \gamma }_\ell }\left( {{m_{{s_\ell }}} - 1} \right)}}} \right)}^{{\zeta _\ell }}}} \right\}} }\notag\\
	&\times\frac{{\Gamma \left( {1 - {\zeta _{L + 1}}} \right)\Gamma \left( {{\zeta _{L + 1}}} \right)}}{{\Gamma \left( {{\zeta _{L + 1}} + 1} \right)}}{s^{{\zeta _{L + 1}}}}d{\zeta _1}d{\zeta _2} \cdots d{\zeta _{L + 1}}.
	\end{align}
	Equation \eqref{orafinal} is obtained using \eqref{zuihouora} and \eqref{definitionH}, which completes the proof.
	\end{IEEEproof}
	\begin{rem}
		With the help of Theorem \ref{P2}, channel capacity under ORA has a tight approximation which can be derived using single $\mathcal{F}$ fading channel. Substituting \eqref{L2ora} into \eqref{orachushi} and letting $L=1$, we get the same result as \cite[eq. (19)]{zhao2019ergodic}.
		An approximation of \cite[eq. (19)]{zhao2019ergodic} in the high SNR region is given by \cite[eq. (26)]{yoo2019comprehensive}. Using \eqref{MatchParematers}, we get a simple algebraic representation of the approximation of channel capacity under ORA in the high SNR region as \cite[eq. (26)]{yoo2019comprehensive}, which also provides useful insights on the impact of the involved parameters. For example, it is evident \cite[eq. (26)]{yoo2019comprehensive} can be expressed in terms of $\overline{\gamma_{{\cal F}}}$. This transformation is useful in quantifying the average SNR value under different fading conditions.

	\end{rem}
	\subsubsection{Optimal Power and Rate Adaptation}
	Under OPRA, the channel capacity is given by \cite[eq. (7)]{goldsmith1997capacity}
	\begin{equation}\label{oprajifen}
	C_{\rm{OPRA}}= B\int_{{\gamma _0}}^\infty  {{{\log }_2}\left( {\frac{z}{{{\gamma _0}}}} \right){f_Z}\left( z \right)} dz
	\end{equation}
	where $\gamma_0$ is the cutoff carrier-to-noise ratio value. No data is sent below $\gamma_0$ and $\gamma_0$ must satisfy \cite[eq. (6)]{goldsmith1997capacity}
	\begin{equation}\label{gamma0}
	\int_{{\gamma _0}}^\infty  {\left( {\frac{1}{{{\gamma _0}}} - \frac{1}{z}} \right)} {f_Z}\left( z \right)dz = 1.
	\end{equation}
	Let
	\begin{equation}
S(z) = \left\{ {\begin{array}{*{20}{l}}
	{\frac{1}{{{\gamma _0}}} - \frac{1}{z}}&{z \ge {\gamma _0}},\\
	0&{{\rm{ otherwise }}},
	\end{array}} \right.
	\end{equation}
	the channel capacity under OPRA can also be written as
	\begin{equation}
	C_{\rm{OPRA}}=B\int_0^\infty  {{{\log }_2}} \left( {1 + S\left( z \right)z} \right){f_Z}\left( z \right)dz.
	\end{equation}
	\begin{prop}
The channel capacity under OPRA is derived as
\begin{equation}\label{oprafinal}
C_{\rm{OPRA}} = \prod\limits_{\ell  = 1}^L {\frac{1}{{\Gamma \left( {{m_\ell }} \right)\Gamma \left( {{m_{{s_\ell }}}} \right)}}} {H_{\rm{OPRA}}}
\end{equation}
where
\begin{align}
{H_{\rm{OPRA}}}\triangleq H_{4,1:2,1; \cdots ;2,1;0,1}^{0,2:1,2; \cdots ;1,2;1,0}\Bigg({\Delta _{\rm{OPRA}}}& \left| {\begin{array}{*{20}{c}}
	{\left( {1; - 1, \cdots , - 1} \right),\left( {1; - 1, \cdots , - 1} \right),\left( {1;1, \cdots ,1} \right),\left( {1;1, \cdots ,1} \right):}\\
	{\left( {1;1, \cdots ,1} \right):}
	\end{array}} \right.\notag\\
& \begin{array}{*{20}{c}}
{\left( {1,1} \right),\left( {1 - {m_{{s_1}}},1} \right); \cdots ;\left( {1,1} \right),\left( {1 - {m_{{s_L}}},1} \right); - }\\
{\left( {{m_1},1} \right); \cdots ;\left( {{m_L},1} \right);\left( {0,1} \right)}
\end{array}\Bigg)
\end{align}
and ${\Delta _{\rm{OPRA}}} \triangleq {\left( {\frac{{{\gamma _0}{m_1}}}{{\left( {{m_{{s_1}}} - 1} \right){{\bar \gamma }_1}}}, \cdots ,\frac{{{\gamma _0}{m_L}}}{{\left( {{m_{{s_L}}} - 1} \right){{\bar \gamma }_L}}},s} \right)^{\rm T}}$.
	\end{prop}
	
	\begin{IEEEproof}
Substituting \eqref{PDFfinal} into \eqref{oprajifen} and changing the order of integration, we can rewrite the channel capacity under OPRA as
\begin{align}\label{oprachushi}
C_{\rm{OPRA} }= &\prod\limits_{\ell  = 1}^L {\frac{1}{{\Gamma \left( {{m_\ell }} \right)\Gamma \left( {{m_{{s_\ell }}}} \right)}}} {\left( {\frac{1}{{2\pi i}}} \right)^L}\int_{{{\cal L}_1}} {\int_{{{\cal L}_2}} { \cdots \int_{{{\cal L}_{_L}}} {\frac{1}{{\Gamma \left( {\sum\limits_{\ell  = 1}^L {{\zeta _\ell }} } \right)}}} \prod\limits_{\ell  = 1}^L {\Upsilon \left( {{\zeta _\ell }} \right)} {{\left( {\frac{{{m_\ell }}}{{{{\bar \gamma }_\ell }\left( {{m_{{s_\ell }}} - 1} \right)}}} \right)}^{{\zeta _\ell }}}}} \notag\\
&\times \underbrace {\int_{{\gamma _0}}^\infty  {{{\log }_2}\left( {\frac{z}{{{\gamma _0}}}} \right){z^{ - 1 + \sum\limits_{\ell  = 1}^L {{\zeta _\ell }} }}dz} }_{{I_{7}}}d{\zeta _1}d{\zeta _2} \cdots d{\zeta _L}.
\end{align}
Let $t = z/{\gamma _0} - 1$ and we have
\begin{equation}
{I_{7}} = {\gamma _0}^{\sum\limits_{\ell  = 1}^L {{\zeta _\ell }} }\int_0^\infty  {{{\log }_2}\left( {t + 1} \right){{\left( {t + 1} \right)}^{ - 1 + \sum\limits_{\ell  = 1}^L {{\zeta _\ell }} }}dt}.
\end{equation}
Using \cite[eq. (2.6.10.49)]{Prudnikov1986Integrals}, \cite[eq. ( \uppercase\expandafter{\romannumeral2}.2)]{Prudnikov1986Integrals} and \cite[eq. (8.331.1)]{gradshteyn2007}, we can solve $I_7$ as
\begin{equation}\label{L2opra}
{I_{7}} = \frac{1}{{\ln 2}}\frac{{\Gamma \left( { - \sum\limits_{\ell  = 1}^L {{\zeta _\ell }} } \right)\Gamma \left( { - \sum\limits_{\ell  = 1}^L {{\zeta _\ell }} } \right){\gamma _0}^{\sum\limits_{\ell  = 1}^L {{\zeta _\ell }} }}}{{\Gamma \left( {1 - \sum\limits_{\ell  = 1}^L {{\zeta _\ell }} } \right)\Gamma \left( {1 - \sum\limits_{\ell  = 1}^L {{\zeta _\ell }} } \right)}}.
\end{equation}
However, eq. \eqref{L2opra} has conflict with the defination of the multivariate Fox's $H$-function, so we choose another way to solve $I_7$. Using the final value theorem and following the similar procedures as the proof of \eqref{cifrfinal}, we can rewrite $I_7$ as
\begin{equation}\label{I7final}
{I_{7}} = \frac{1}{{2\pi j}}{\gamma _0}^{\sum\limits_{\ell  = 1}^L {{\zeta _\ell }} }\int_{{{\cal L}_{L + 1}}} {\frac{{{\Gamma ^2}\left( { - \sum\limits_{\ell  = 1}^{L + 1} {{\zeta _\ell }} } \right)\Gamma \left( { - {\zeta _{L + 1}}} \right)}}{{{\Gamma ^2}\left( {1 - \sum\limits_{\ell  = 1}^{L + 1} {{\zeta _\ell }} } \right)}}{s^{{\zeta _{L + 1}}}}} d{\zeta _{L + 1}}.
\end{equation}
Substituting \eqref{I7final} into \eqref{oprachushi}, we obtain
\begin{align}
 C_{\rm{OPRA}}= & \prod\limits_{\ell  = 1}^L {\frac{1}{{\Gamma \left( {{m_\ell }} \right)\Gamma \left( {{m_{{s_\ell }}}} \right)}}} {\left( {\frac{1}{{2\pi i}}} \right)^{L + 1}}\int_{{{\cal L}_1}}\! {\int_{{{\cal L}_2}} {\! \!\cdots\! \!\int_{{{\cal L}_{_{L + 1}}}}\!}{\frac{{{\Gamma ^2}\left( { - \sum\limits_{\ell  = 1}^{L + 1} {{\zeta _\ell }} } \right)}}{{\Gamma \left( {\sum\limits_{\ell  = 1}^L {{\xi _\ell }} } \right){\Gamma ^2}\left( {1 - \sum\limits_{\ell  = 1}^{L + 1} {{\zeta _\ell }} } \right)}}}}\notag\\
& \!\times\!\! \prod\limits_{\ell  = 1}^L {\Upsilon \left( {{\zeta _\ell }} \right)} {{\left( {\frac{{{\gamma _0}{m_\ell }}}{{{{\bar \gamma }_\ell }\left( {{m_{{s_\ell }}} - 1} \right)}}} \right)}^{{\zeta _\ell }}} \Gamma \left( { - {\zeta _{L + 1}}} \right){s^{{\zeta _{L + 1}}}}d{\zeta _1}d{\zeta _2} \cdots d{\zeta _{L + 1}}.
\end{align}
With the help of \eqref{definitionH}, we obtain \eqref{oprafinal}. The proof is now complete.
	\end{IEEEproof}
	\begin{rem}
	Based on Theorem \ref{P2}, A tight approximation of channel capacity under OPRA can be derived using single $\mathcal{F}$ fading channel. Substituting \eqref{L2opra} into \eqref{oprachushi} and letting $L=1$, the same result as \cite[eq. (52)]{yoo2019comprehensive} is obtained as
	Note that \cite[eq. (52)]{yoo2019comprehensive} is suitable for analysis both analytically and numerically and can used as a benchmark for further derivation of an approximation.	
	In the high-SNR regime, the approximation of \eqref{oprafinal} is given as \cite[eq. (57)]{yoo2019comprehensive}.
	\end{rem}
	\begin{rem}
		In order to determine the value of $\gamma_0$, we can rewrite \eqref{gamma0} as
		\begin{equation}
		\frac{1}{{{\gamma _0}}}\underbrace {\int_{{\gamma _0}}^\infty  {{f_Z}} (z)dz}_{{I_8}} - \underbrace {\int_{{\gamma _0}}^\infty  {\frac{{{f_Z}(z)}}{z}} dz}_{{I_9}} = 0
		\end{equation}
		where $I_8$ and  $I_9$ have been solved in \eqref{ccdftocdf} and \eqref{tifrfinaljifen} respectively. Thus, it follows that
		\begin{equation}\label{ga0}
		\frac{1}{{{\gamma _0}}}(1-\prod\limits_{\ell  = 1}^L {\frac{1}{{\Gamma \left( {{m_{{s_\ell }}}} \right)\Gamma \left( {{m_\ell }} \right)}}} {H_{{\rm{CDF}}}}) -  {{\prod\limits_{\ell  = 1}^L {\frac{1}{{\Gamma \left( {{m_{{s_\ell }}}} \right)\Gamma \left( {{m_\ell }} \right)}}} {H_{{\rm{TIFR}}}}}}= 0.
		\end{equation}
		A detailed proof of the existence of $\gamma_0$ in the range $\left[ {0,1} \right]$ is given in \cite{cheng2001capacity}. Thus, with the aid of \eqref{ga0}, $\gamma_0$ can be solved with mathematical software and
		we find that using the dichotomy for iterations within 20 times can make the error less than ${10}^{-6}$.
	\end{rem}
	\begin{rem}\label{rem7}
	Channel capacity in AWGN, in bits per second, is given by Shannon’s formula as \cite{goldsmith2005wireless}
	\begin{equation}
	C_{\rm{AWGN}}=B \log _{2}(1+z).
	\end{equation}
	The relationship of $C_{\rm{ORA}}$, $C_{\rm{OPRA}}$ and $C_{\rm{AWGN}}$  can be obtained by applying the Jensen's inequality as \cite{alouini2000comparison}
	\begin{subequations}
	\begin{equation}
	0 \le C_{\rm{OPRA}}-C_{\rm{ORA}} \le \min \left(C_{\rm{OPRA}},-\log _{2} \gamma_{0}\right)
	\end{equation}
	\begin{equation}
	C_{\rm{ORA} }\le C_{\rm{AWGN}}
	\end{equation}
	\end{subequations}
	and there is no general order relation between $C_{\rm{OPRA}}$ and $C_{\rm{AWGN}}$.
	\end{rem}

	\section{Numerical Results}\label{Section6}
	In this section, analytical results are presented to illustrate the proposed applications of the sum of Fisher-Snedecor $\mathcal{F}$ RVs in wireless communication systems. All results are substantiated by Monte Carlo simulations.
	\begin{figure}[t]
	\centering
	\includegraphics[scale=1]{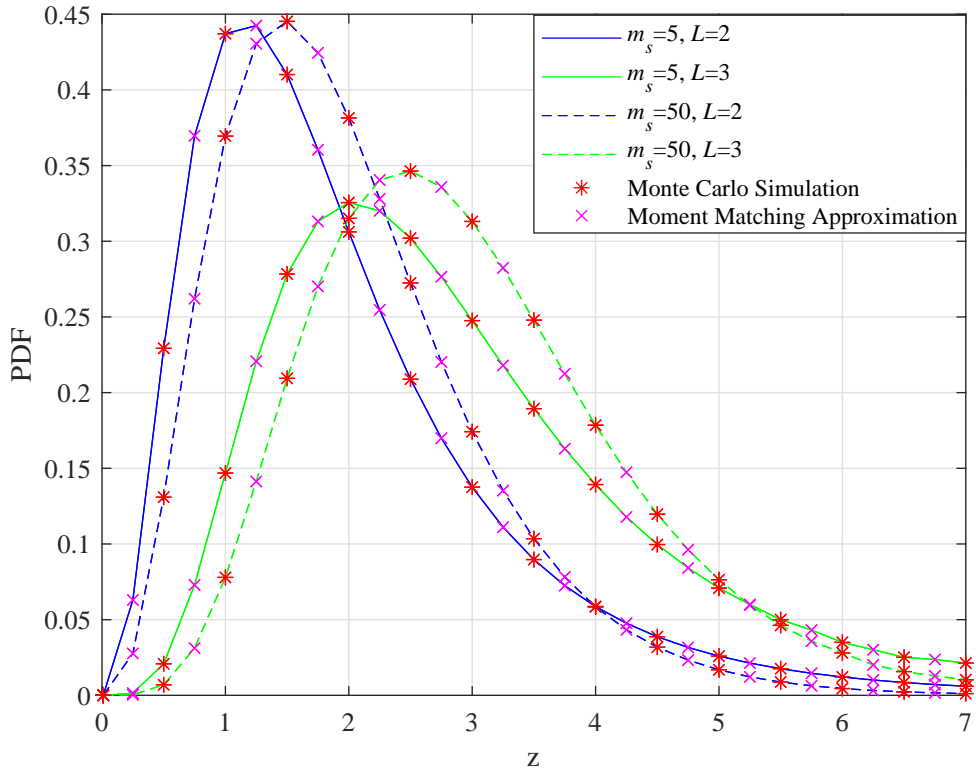}
	\caption{PDF for the sum of squared $\mathcal{F}$-distributed RVs.}
	\label{sumPDF}
	\end{figure}
	\begin{figure}[t]
	\centering
	\includegraphics[scale=1]{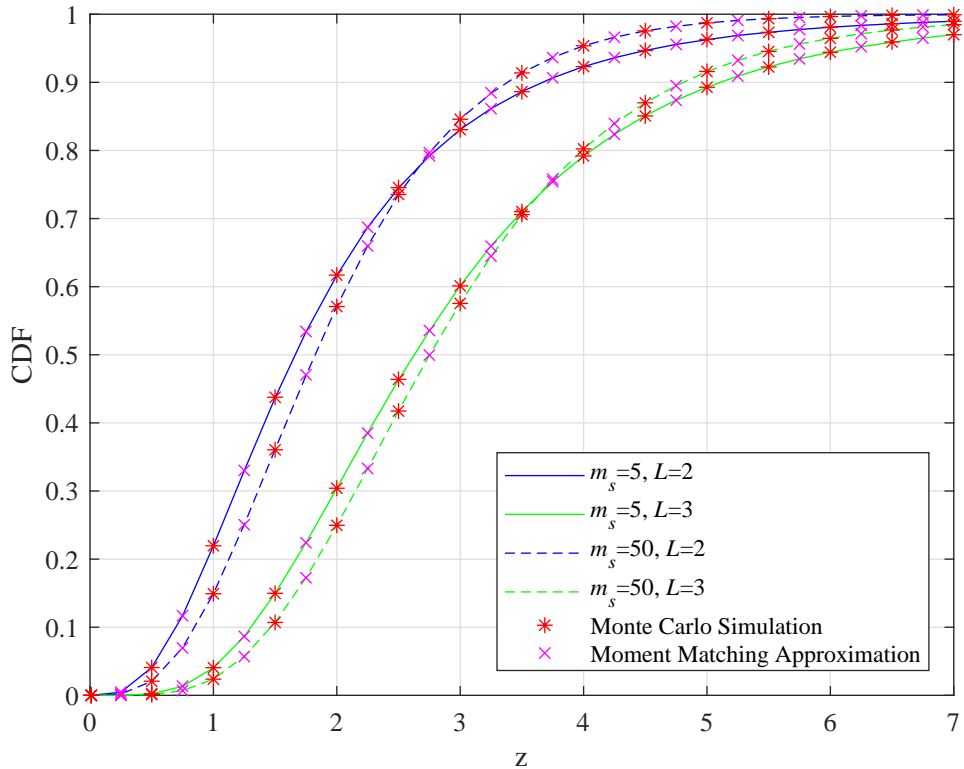}
	\caption{CDF for the sum of squared $\mathcal{F}$-distributed RVs.}
	\label{sumCDF}
	\end{figure}
	
Figures \ref{sumPDF} and \ref{sumCDF}, respectively, plot the PDF and CDF of the sum of Fisher-Snedecor $\mathcal{F}$ RVs and their proposed approximation obtained by moment matching method for different values of $m_s$ and $L$, assuming $m_{s_\ell}=m_{s}$ $(\ell=1,2,3)$, $m_\ell=2$, $\gamma_\ell=0$ ${\rm{dB}}$.  In both figures, it can be observed that the approximate PDFs and CDFs match the exact ones well for all considered cases. Furthermore, analytical results perfectly match Monte Carlo simulations, thus validating our results.
	\begin{figure}[t]
	\centering
	\includegraphics[scale=1]{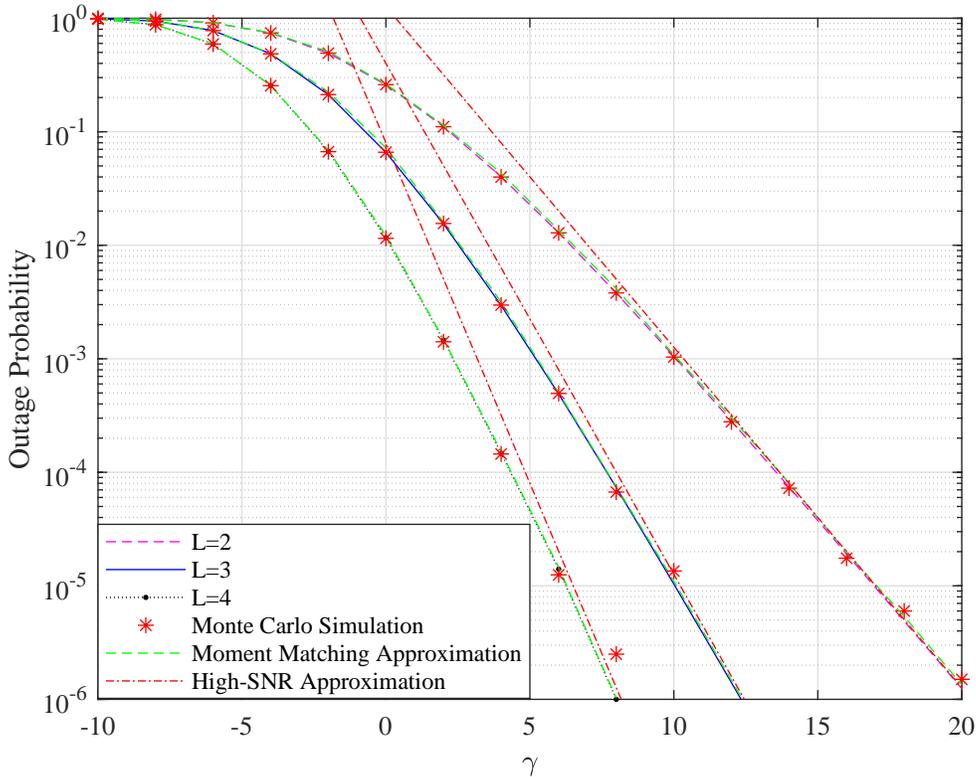}
	\caption{Outage probability versus average SNR for dual, triple and quadruple MRC receivers.}
	\label{OPeps}
	\end{figure}
	
	Figure \ref{OPeps} depicts the OP performance versus average SNR $\gamma$ for a dual-, triple- and quadruple-branch	MRC receivers with $\gamma_\ell=\gamma$ $(\ell=1,2,3,4)$, $\gamma_{\rm{th}}=0$ ${\rm{dB}}$, $m_\ell=1.5$, $m_{s_\ell}=5$. As it can be observed, the OP decreases as the average SNR and $L$ increase. Again, it is evident that the exact results match the approximate ones and Monte Carlo simulations well. In addition, the asymptotic expressions match well the exact ones at high-SNR values thus proving their validity and versatility. The OP performance improvement is more pronounced by increasing $L=2$ to $L=3$.
	\begin{figure}[t]
	\centering
	\includegraphics[scale=1]{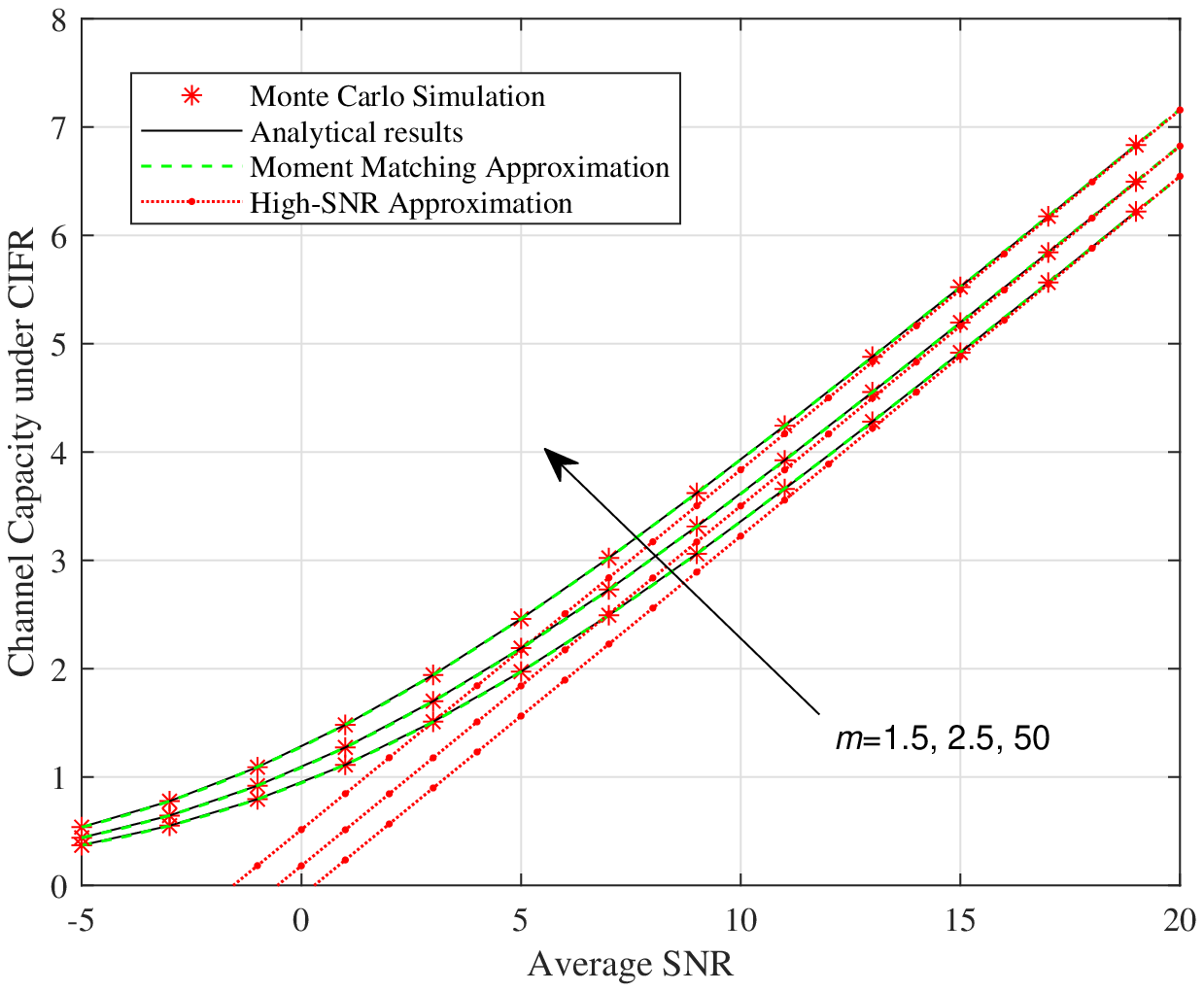}
	\caption{Channel capacity under CIFR versus average SNR for L=2.}
	\label{cifreps}
	\end{figure}
	\begin{figure}[t]
	\centering
	\includegraphics[scale=1]{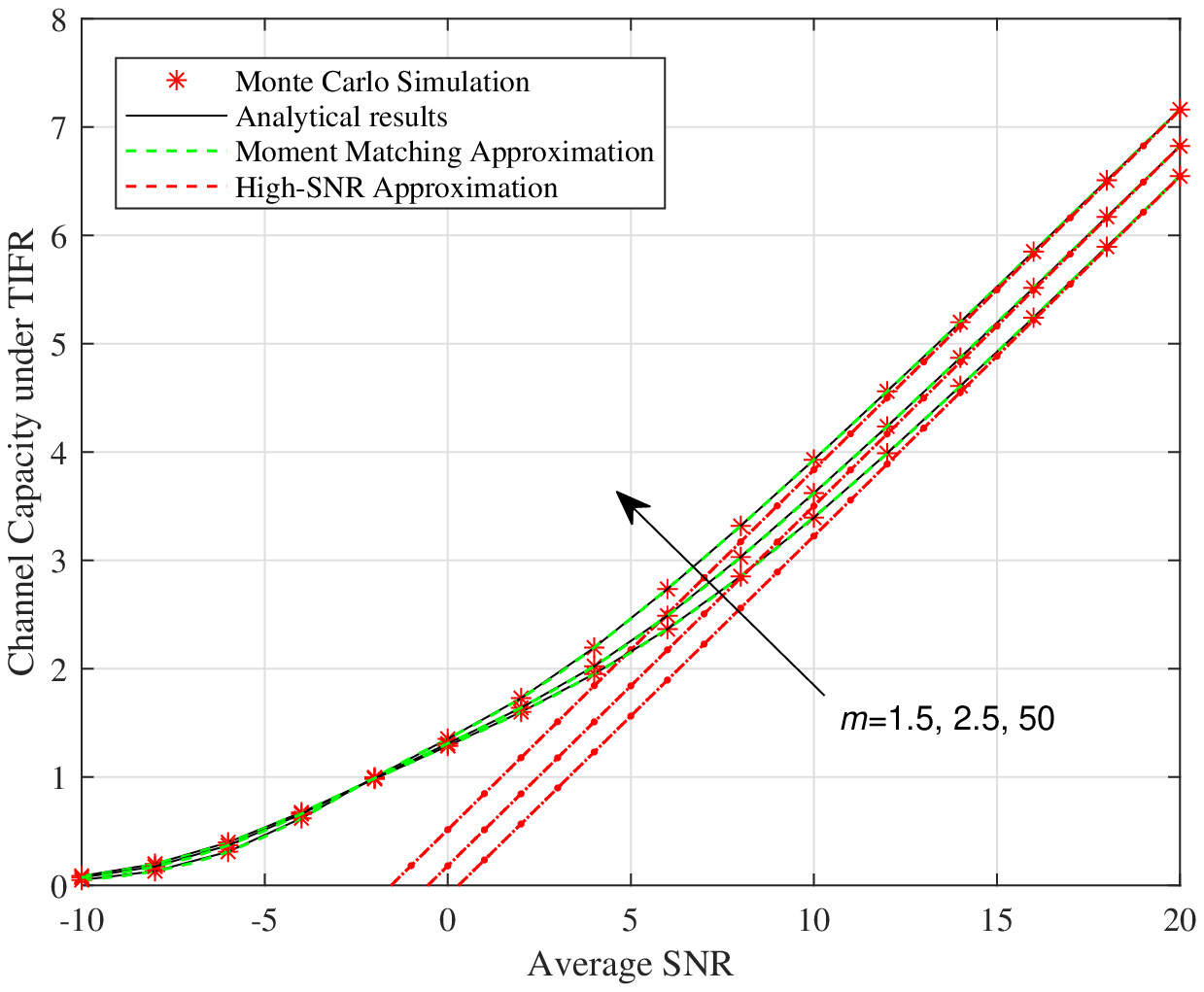}
	\caption{Channel capacity under TIFR versus average SNR for L=2.}
	\label{tifreps}
	\end{figure}
	\begin{figure}[t]
	\centering
	\includegraphics[scale=1]{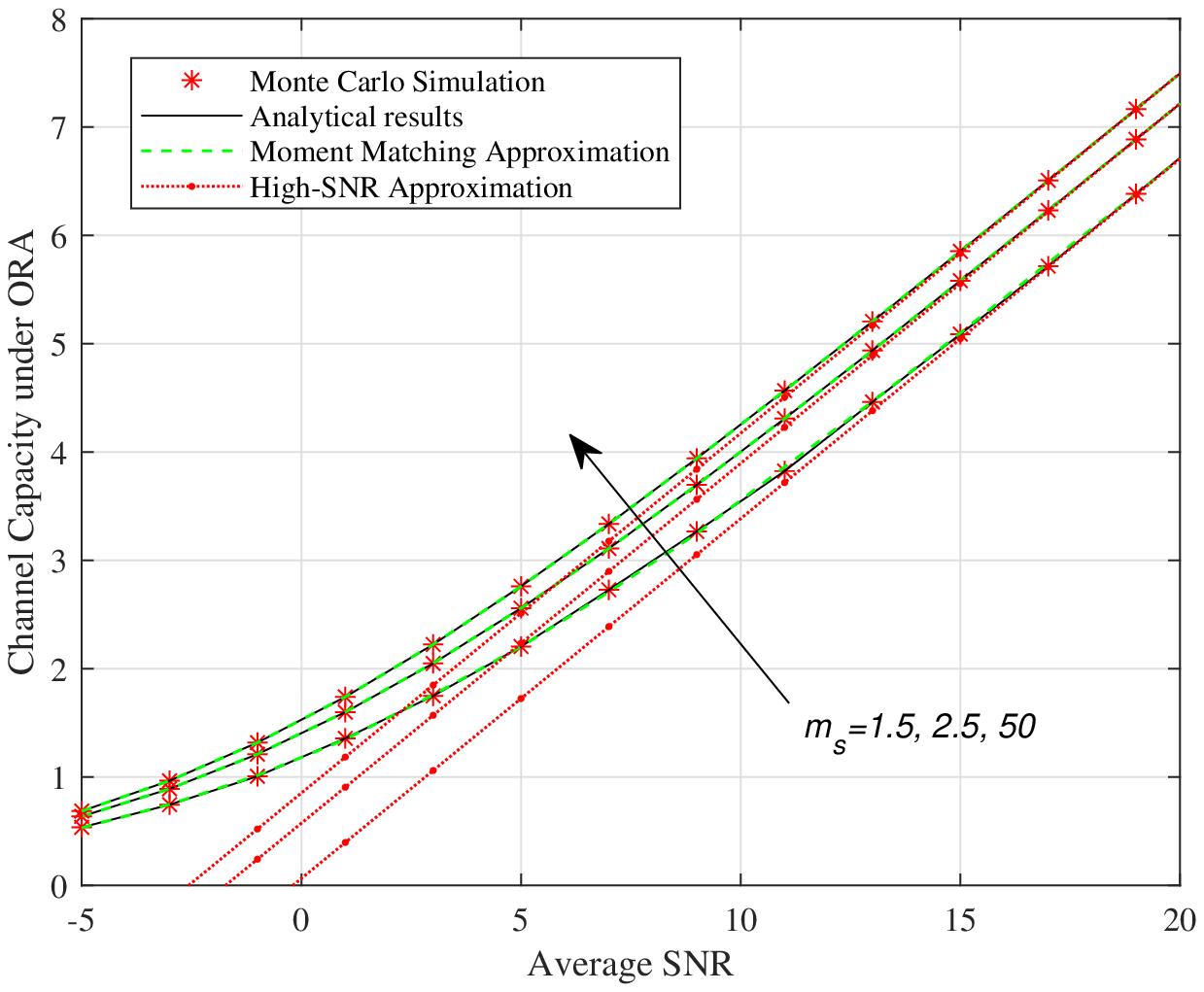}
	\caption{Channel capacity under ORA versus average SNR for L=2.}
	\label{oraeps}
	\end{figure}
	\begin{figure}[t]
	\centering
	\includegraphics[scale=1]{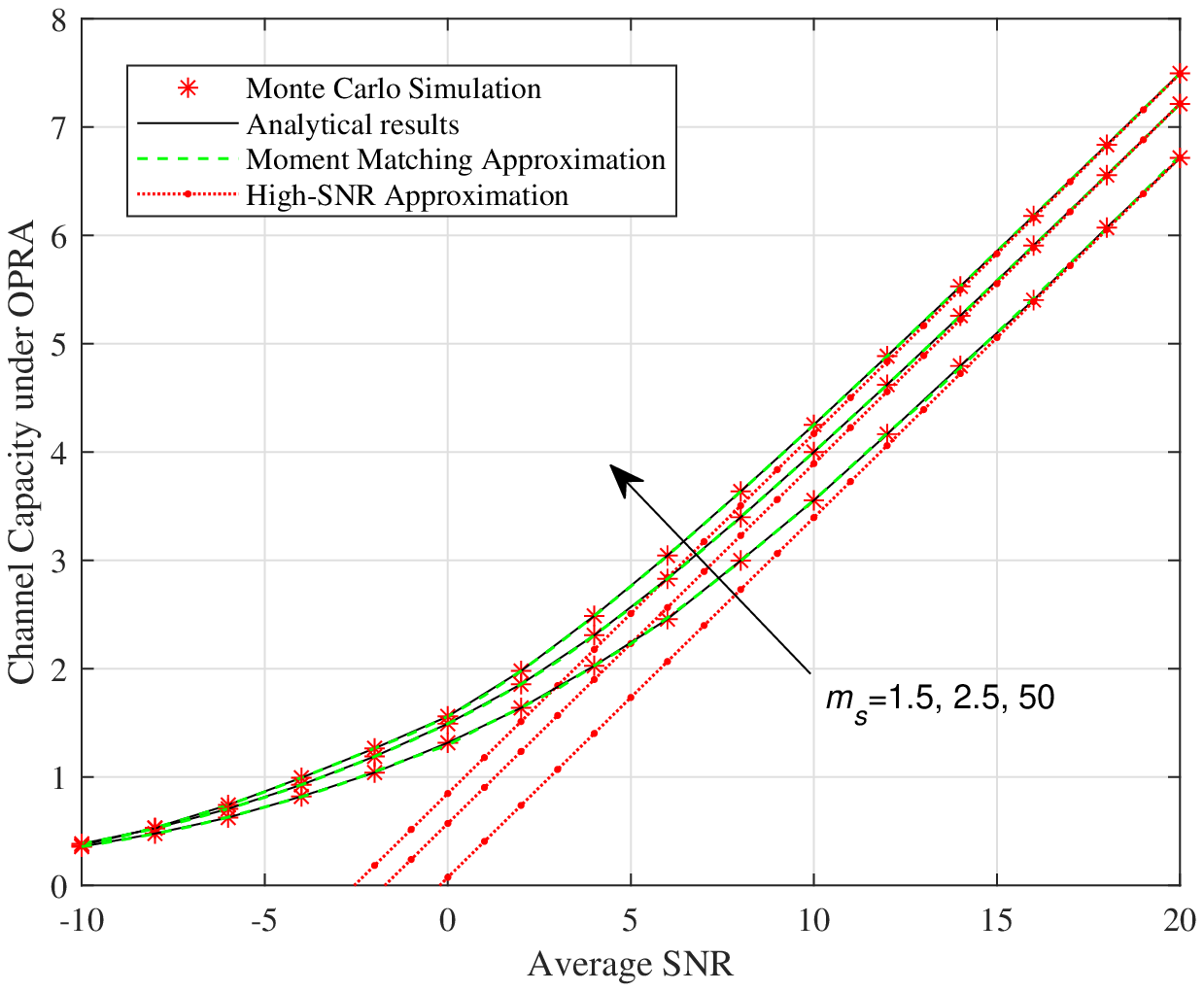}
	\caption{Channel capacity under OORA versus average SNR for L=2.}
	\label{opraeps}
	\end{figure}
	
	Figures \ref{cifreps}-\ref{opraeps}, respectively, show the analytical and simulated channel capacities versus average SNR $\gamma$ under different adaptive transmission strategies respectively, assuming $L=2$. Figs. \ref{cifreps} and \ref{tifreps} illustrate the channel capacity increases as $m$ increases, while Figs. \ref{oraeps} and \ref{opraeps} depict the channel capacity increases as $m_s$ increases, which means favorable system parameters can lead to a large channel capacity. Once again, perfect agreement is observed between analytical results, approximate results and Monte Carlo simulations in Figs. \ref{cifreps}-\ref{opraeps}.
	\begin{figure}[t]
	\centering
	\includegraphics[scale=1]{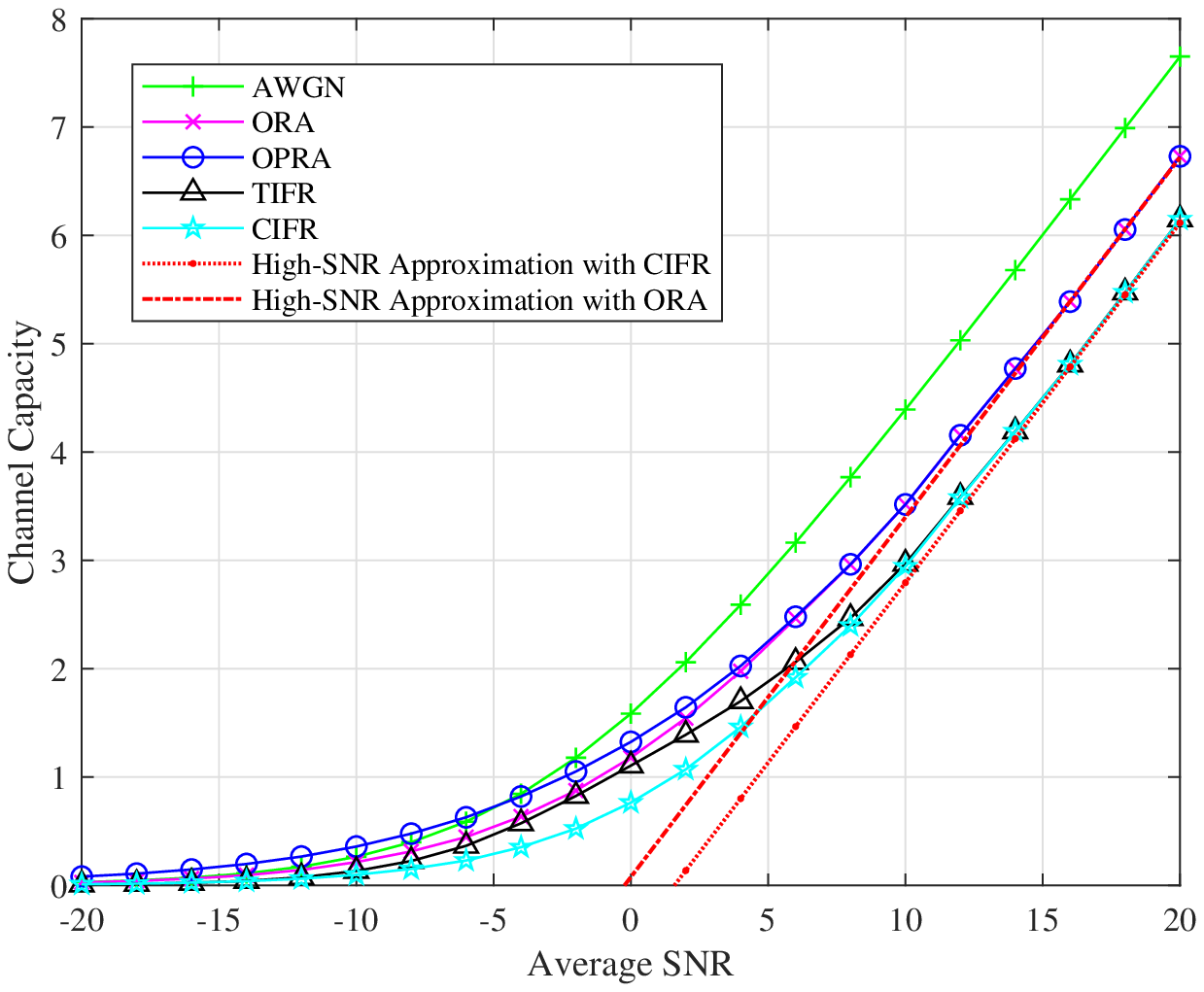}
	\caption{Channel capacity under four different adaptive transmission strategies versus average SNR for L=2.}
	\label{duibi}
	\end{figure}
	
	As shown in Fig. \ref{duibi}, four different adaptive transmission strategies provide different channel capacities. Obviously, the relation we proposed in \eqref{rem7} is shown. As it can be observed, the channel capacity under OPRA is the largest among those adaptive transmission strategies, followed by the channel capacity under ORA. The channel capacity under TIFR and CIFR converges in the high-SNR regime because the asymptotic expressions of those two cases are the same for $m>1$. This finding can be also observed in channel capacity under OPRA and ORA. Moreover, analytical and approximate results agree well with Monte Carlo simulations.
	
	\begin{figure}[t]
	\centering
	\includegraphics[scale=1]{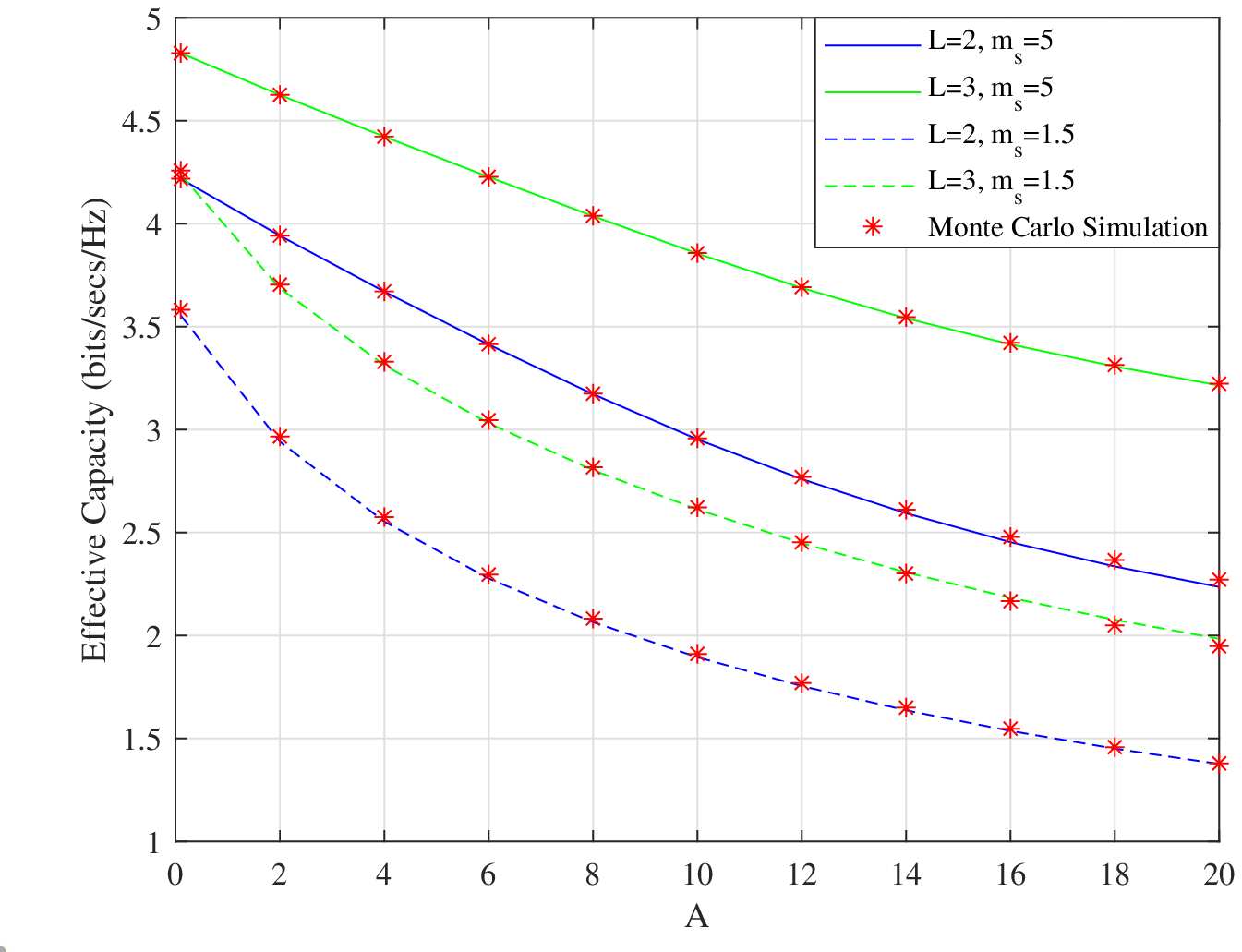}
	\caption{Effective capacity of MRC receivers over i.n.i.d. $\mathcal{F}$ fading channels versus the delay constraint $A$.}
	\label{effeps}
	\end{figure}
		Figure \ref{effeps} illustrates the effective capacity of MRC systems over i.n.i.d. $\mathcal{F}$ fading channels as a function of the delay constraint $A$ for different settings of the parameters $m_s$ and $L$, assuming $m_{s_\ell}=m_s$, $m_\ell=5$, $\gamma_\ell=10$ ${\rm{dB}}$ $(\ell=1,2,3)$. It can be easily observed that, the effective capacity increases as $m_s$ and $L$ increase. This is because of the fact that large value of $m_s$ induces low shadowing effect and large value of $L$ means more receive power. For a certain setting of the parameters, the effective capacity increases as $A$ decreases (i.e. large delay). Thus, the considered system can support larger arrival rates
		with a larger delay constraint. Fig. \ref{effeps} shows a perfect agreement of the Monte Carlo simulations and analytical and simulation results, which confirms the validity of the analytical result in \eqref{efffinal}.
	\section{Conclusion}\label{Section7}
	We presented exact expressions for the PDF and CDF of i.n.i.d. Fisher-Snedecor $\mathcal{F}$ RVs. In addtion, we proposed the single $\mathcal{F}$ distribution with an adjusted form of parameters to approximate the sum of $\mathcal{F}$ distribution by using the moment matching method. The derived results show that the introduced adjustment results closely approximate in both the lower and upper tail regions. This sufficiently accurate region-wise approximation significantly simplify the performance analysis of MRC diversity	receivers over $\mathcal{F}$ fading channels. To this end, novel, computationally efficient analytical expressions were obtained for OP, effective capacity and channel capacities under different adaptive transmission strategies. Finally, Extensive numerical results have been presented to validate the proposed analytical expressions and an excellent agreement has been observed. The new and important insights we provided are useful for wireless communication system designers.
	
	\begin{appendices}
	\section{Proof of Theorem \ref{P1}}\label{AppendixA}
	\renewcommand{\theequation}{A-\arabic{equation}}
	\setcounter{equation}{0}
	The PDF of $z$ can be obtained as
	\begin{equation}\label{LP}
	{f_z }\left( z\right) = {{\cal L}^{{-1}}}\left\{ {{{\cal M}_z }\left( s \right);z} \right\} = \frac{1}{{2\pi j}}\int_{\cal L} {{{\cal M}_z }\left( s \right){e^{z s}}} ds
	\end{equation}
	where ${{\cal L}^{{-1}}}\left\{{\cdot}\right\}$ denotes the inverse Laplace transform and ${{\cal M}_z}\left(s\right)$ is the MGF of $z$.
	Because the Fisher-Snedecor $\mathcal{F}$ RVs are independent, we can obtain the MGF of the $z$ using \eqref{MGF_F2} as
	\begin{equation}\label{MGFz}
	{{\cal M}_z}\left( s \right) = \prod\limits_{\ell  = 1}^L {{{\cal M}_{{\gamma _\ell }}}\left( s \right)}  = \prod\limits_{\ell  = 1}^L {\frac{{\Gamma \left( {{m_\ell } + {m_{{s_\ell }}}} \right)}}{{\Gamma \left( {{m_{{s_\ell }}}} \right)}}\Psi \left( {{m_\ell };1 - {m_{{s_\ell }}};\frac{{s{{\bar \gamma }_\ell }\left( {{m_{{s_\ell }}} - 1} \right)}}{{{m_\ell }}}} \right)}.
	\end{equation}
	Substituting \eqref{MGFz} into \eqref{LP} and using \cite[eq. (8.4.46.1)]{prudnikov2003integrals}, we can derive the MGF of $z$ as
	\begin{equation}\label{MGFcal}
	{f_z}\left( z \right) = \frac{1}{{2\pi j}}\int_{\cal L} {\prod\limits_{\ell  = 1}^L {\frac{1}{{\Gamma \left( {{m_{{s_\ell }}}} \right)\Gamma \left( {{m_\ell }} \right)}}G_{1,2}^{2,1}\left( {\left. {\frac{{s{{\bar \gamma }_\ell }\left( {{m_{{s_\ell }}} - 1} \right)}}{{{m_\ell }}}} \right|\begin{array}{*{20}{c}}
	{1 - {m_\ell }}\\
	{0,{m_{{s_\ell }}}}
	\end{array}} \right)} {e^{zs}}} ds
	\end{equation}
	where the integration path of  ${\cal L}$ goes from $\sigma -\infty j$ to $\sigma+\infty j$ and $\sigma  \in \mathbb{R}$. Rewriting each Meijer’s $G$-function as a Mellin-Barnes integral with the aid of \cite[eq. (8.4.3.1)]{gradshteyn2007} and merging the integrals, we obtain
	\begin{align}\label{ML+1}
	{f_z}\left( z \right) =&\prod\limits_{\ell  = 1}^L {\frac{1}{{\Gamma \left( {{m_{{s_\ell }}}} \right)\Gamma \left( {{m_\ell }} \right)}}\int_{{{\cal L}_1}} {\int_{{{\cal L}_2}} { \cdots \int_{{{\cal L}_{_L}}} {{{\left( {\frac{1}{{2\pi j}}} \right)}^L}\prod\limits_{\ell  = 1}^L {\Gamma \left( { - {t_\ell }} \right)\Gamma \left( {{m_\ell } + {t_\ell }} \right)\Gamma \left( {{m_\ell } + {m_{{s_\ell }}} + {t_\ell }} \right)} } } } } \notag\\
	&\times {\left( {\frac{{{{\bar \gamma }_\ell }\left( {{m_{{s_\ell }}} - 1} \right)}}{{{m_\ell }}}} \right)^{ - \left( {{m_\ell } + {t_\ell }} \right)}}\underbrace {\frac{1}{{2\pi j}}\int_{\cal L} {{s^{ - \sum\limits_{\ell  = 1}^L {{m_\ell }}  - \sum\limits_{\ell  = 1}^L {{t_\ell }} }}} {e^{zs}}ds}_{{I_{{A}}}}d{t_1}d{t_2} \cdots d{t_L}
	\end{align}
	where ${\cal L}_\ell$ $(\ell=1, \cdots, L)$, goes from $\sigma_\ell -\infty j$ to $\sigma_\ell+\infty j$ and $\sigma_\ell \in \mathbb{R}$. Using \cite[eq. (8.315.1)]{gradshteyn2007}, we can solve ${I_{A}}$ as
	\begin{equation}\label{IA1}
	{I_{{A}}} = {z^{ - 1 + \sum\limits_{\ell  = 1}^L {{m_\ell }}  + \sum\limits_{\ell  = 1}^L {{t_\ell }} }}\frac{1}{{\Gamma \left( {\sum\limits_{\ell  = 1}^L {{m_\ell }}  + \sum\limits_{\ell  = 1}^L {{t_\ell }} } \right)}}.
	\end{equation}
	Let ${\zeta_\ell} \triangleq {m_\ell}+{t_\ell}$. Substitute \eqref{IA1} into \eqref{ML+1}, we can write \eqref{ML+1} as
	\begin{align}\label{MGFofZ}
	{f_z}\left( z \right) =&\frac{1}{z}\prod\limits_{\ell  = 1}^L {\frac{1}{{\Gamma \left( {{m_{{s_\ell }}}} \right)\Gamma \left( {{m_\ell }} \right)}}\int_{{{\cal L}_1}} {\int_{{{\cal L}_2}} { \cdots \int_{{{\cal L}_{_L}}} {{{\left( {\frac{1}{{2\pi j}}} \right)}^L}\frac{1}{{\Gamma \left( {\sum\limits_{\ell  = 1}^L {{\zeta _\ell }} } \right)}}} } } } \notag\\
	& \times \prod\limits_{\ell  = 1}^L {\Gamma \left( {{m_\ell } - {\zeta _\ell }} \right)\Gamma \left( {{\zeta _\ell }} \right)\Gamma \left( {{m_{{s_\ell }}} + {\zeta _\ell }} \right)} {\left( {\frac{{z{m_\ell }}}{{{{\bar \gamma }_\ell }\left( {{m_{{s_\ell }}} - 1} \right)}}} \right)^{  {\zeta _\ell }}}d{\zeta _1}d{\zeta _2} \cdots d{\zeta _L}.
	\end{align}
The proof is completed by deriving \eqref{PDFfinal} with the definition of the multivariate $H$-function \cite[eq. (A.1)]{mathai2009h}.
	\section{Proof of Theorem \ref{P2}}\label{AppendixB}
	\renewcommand{\theequation}{B-\arabic{equation}}
	\setcounter{equation}{0}
	Let us start with $L=2$ and $z=\gamma_1+\gamma_2$, $z \sim \mathcal{F}\left(\overline{\gamma}, m, m_{s}\right)$ and $\gamma_\ell \sim \mathcal{F}\left(\overline{\gamma}_\ell, m_\ell, m_{s_\ell}\right)$ $(\ell=1,2)$. The first, second and third moments of the sum of two independent RVs can be written as
	\begin{equation}\label{2moment3eq}
	\left\{ \begin{array}{l}
	{\rm{E}}[z] = {\rm{E}}[\gamma_1] + {\rm{E}}[\gamma_2],\\
	{\rm E}\left[ {{z^2}} \right] = {\rm E}\left[ {{\gamma_1^2}} \right] + {\rm E}\left[ {{\gamma_2^2}} \right] + 2{\rm E}\left[ \gamma_1 \right]{\rm E}\left[\gamma_2 \right],\\
	{\rm E}\left[ {{z^3}} \right] = {\rm E}\left[ {{\gamma_1^3}} \right] + {\rm E}\left[ {{\gamma_2^3}} \right] + 3{\rm E}\left[ {{\gamma_1^2}} \right]{\rm E}\left[\gamma_2 \right] + 3{\rm E}\left[ \gamma_1 \right]{\rm E}\left[ {{\gamma_2^2}} \right],
	\end{array} \right.
	\end{equation}
With the help of \eqref{momenteq}, eq. \eqref{2moment3eq} can be written as
	\begin{equation}\label{3moment3eq}
	\left\{ \begin{array}{l}
	\bar \gamma_{\cal F} {=}{{\bar \gamma }_{1}} + {{\bar \gamma }_{2}},\\
	{H_{\cal F}}{{\bar \gamma }^2} = {H_{1}}{{\bar \gamma }_1}^2 + {H_2}{{\bar \gamma }_2}^2{+2}{{\bar \gamma }_1}{{\bar \gamma }_2},\\
	{H_{\cal F}}{Y_{\cal F}}{{\bar \gamma }^3} = {H_1}{Y_1}{{\bar \gamma }_1}^3 + {H_2}{Y_2}{{\bar \gamma }_2}^3 + 3{H_1}{{\bar \gamma }_1}^2{{\bar \gamma }_2} + 3{H_2}{{\bar \gamma }_1}{{\bar \gamma }_2}^2
	\end{array} \right.
	\end{equation}
	where $H_{\cal F}=\frac{{\left( {1 + m} \right)\left( {{m_s} - 1} \right)}}{{m\left( {{m_s} - 2} \right)}}$, $Y_{\cal F} = \frac{{\left( {{m_s} - 1} \right)\left( {2 + m} \right)}}{{m\left( {{m_s} - 3} \right)}}$, ${H_\ell } = \frac{{\left( {1 + {m_\ell }} \right)\left( {{m_{{s_\ell }}} - 1} \right)}}{{{m_\ell }\left( {{m_{{s_\ell }}} - 2} \right)}}$, ${Y_\ell } = \frac{{\left( {{m_{{s_\ell }}} - 1} \right)\left( {2 + {m_\ell }} \right)}}{{{m_\ell }\left( {{m_{{s_\ell }}} - 3} \right)}}$ $(\ell=1,2)$. In the composite fading channel, an $\mathcal{F}$ fading channel's amount of fading (AF), which is often used as a relative measure of the severity of fading, is derived in \cite{7886273} as $\left( {{H_{\cal F}} - 1} \right)$. By solving \eqref{3moment3eq}, we obtain
	
	\begin{equation}\label{solvemoment}
	\left\{ \begin{array}{l}
	\bar \gamma_{\cal F}  = {{\bar \gamma }_1} + {{\bar \gamma }_2},\\
	{H_{\cal F}} = \frac{{{H_1}{{\bar \gamma }_1}^2 + {H_2}{{\bar \gamma }_2}^2 + 2{{\bar \gamma }_1}{{\bar \gamma }_2}}}{{{{\left( {{{\bar \gamma }_1} + {{\bar \gamma }_2}} \right)}^{2}}}} = \frac{{\left( {{H_1} - 1} \right){{\bar \gamma }_1}^2 + \left( {{H_2} - 1} \right){{\bar \gamma }_2}^2}}{{{{\left( {{{\bar \gamma }_1} + {{\bar \gamma }_2}} \right)}^{2}}}} + 1,\\
	{Y_{\cal F}}{=}\frac{{{H_1}{Y_1}{{\bar \gamma }_1}^3 + {H_2}{Y_2}{{\bar \gamma }_2}^3 + 3{{\bar \gamma }_1}^2{{\bar \gamma }_2}{H_1} + 3{{\bar \gamma }_1}{{\bar \gamma }_2}^2{H_2}}}{{\left( {{{\bar \gamma }_1} + {{\bar \gamma }_2}} \right)\left( {{H_1}{{\bar \gamma }_1}^2 + {H_2}{{\bar \gamma }_2}^2 + 2{{\bar \gamma }_1}{{\bar \gamma }_2}} \right)}}\\
	{\kern 12pt}= \frac{{\left( {{H_1}{Y_1} - 1} \right){{\bar \gamma }_1}^3 + \left( {{H_2}{Y_2} - 1} \right){{\bar \gamma }_2}^3 + {{\left( {{{\bar \gamma }_1} + {{\bar \gamma }_2}} \right)}^{3}} + 3\left( {{{\bar \gamma }_1} + {{\bar \gamma }_2}} \right)\left( {{{\bar \gamma }_1}^2\left( {{H_1} - 1} \right) + {{\bar \gamma }_2}^2\left( {{H_2} - 1} \right)} \right) - 3\left( {{{\bar \gamma }_1}^3\left( {{H_1} - 1} \right) + {{\bar \gamma }_2}^3\left( {{H_2} - 1} \right)} \right)}}{{\left( {{{\bar \gamma }_1} + {{\bar \gamma }_2}} \right)\left( {\left( {{H_1} - 1} \right){{\bar \gamma }_1}^2 + \left( {{H_2} - 1} \right){{\bar \gamma }_2}^2 + {{\left( {{{\bar \gamma }_1} + {{\bar \gamma }_2}} \right)}^{2}}} \right)}}.
	\end{array} \right.
	\end{equation}
	Equations \eqref{solvemoment} can be generalized for the sum of $L$ i.n.i.d. Fisher-Snedecor $\mathcal{F}$ RVs as \cite{al2010approximation}
	\begin{equation}\label{withoute}
	\left\{ \begin{array}{l}
	\bar \gamma_{\cal F} {=}\sum\limits_{\ell = 1}^L {{{\bar \gamma }_\ell}},\\
	{H_{\cal F}} = \frac{{\sum\limits_{\ell = 1}^L {\left( {{H_\ell} - 1} \right){{\bar \gamma }_\ell}^2} }}{{{{\left( {\sum\limits_{\ell = 1}^L {{{\bar \gamma }_\ell}} } \right)}^{2}}}} + 1,\\
	{Y_{\cal F}} = \frac{{\sum\limits_{\ell = 1}^L {\left( {{H_\ell}{Y_\ell} - 1} \right){{\bar \gamma }_\ell}^3}  + {{\left( {\sum\limits_{\ell = 1}^L {{{\bar \gamma }_\ell}} } \right)}^3} + 3\left( {\sum\limits_{\ell = 1}^L {{{\bar \gamma }_\ell}} } \right)\left( {\sum\limits_{\ell = 1}^L {\left( {{H_\ell} - 1} \right){{\bar \gamma }_\ell}^2} } \right) - 3\sum\limits_{\ell = 1}^L {\left( {{H_\ell} - 1} \right){{\bar \gamma }_\ell}^3} }}{{\left( {\sum\limits_{\ell = 1}^L {{{\bar \gamma }_\ell}} } \right)\left( {\sum\limits_{\ell = 1}^L {\left( {{H_\ell} - 1} \right){{\bar \gamma }_\ell}^2} {+}{{\left( {\sum\limits_{\ell = 1}^L {{{\bar \gamma }_\ell}} } \right)}^{2}}} \right)}}.
	\end{array} \right.
	\end{equation}
	However, the parameters of a single $\mathcal{F}$ distribution calculated by \eqref{withoute} result in a relatively large approximation error in the lower and upper tail regions because we only match the first, second and third moments. This error can be modified by introducing an adjustment factor $\varepsilon$.
	The proof is complete by deriving \eqref{MatchParematers} and \eqref{MomentMatch}.
	
	\end{appendices}

\newpage
	\bibliographystyle{IEEEtran}
	\bibliography{IEEEabrv,Ref}
	
	\end{document}